\documentclass[aps,prd,11pt,reprint,nofootinbib,showkeys]{revtex4-1} 
\usepackage{amsmath,amssymb,amsthm,mathrsfs,bbm}
\usepackage{latexsym,amscd,amsbsy,amsfonts,dsfont}
\usepackage{graphicx}
\usepackage[utf8]{inputenc}
\usepackage{subfigure}  
\usepackage{float}
\usepackage{slashed}
\usepackage{cancel}
\usepackage{multirow}
\usepackage{soul}
\usepackage{physics}
\usepackage{comment}
\usepackage[usenames,dvipsnames]{xcolor}
\usepackage[
      colorlinks=true,
      linktocpage=true,
      breaklinks=true,
      linkcolor=Aquamarine,
      urlcolor=Purple,
      filecolor=black,
      citecolor=Purple,
      menucolor=black,
      pdfstartview=FitV,
      bookmarksopen=true
      ]{hyperref}

\newcommand\ee{\mathrm{e}}
\newcommand\ii{\mathrm{i}}
\newcommand\dimM{n}
\newcommand\sT{{\scriptscriptstyle \mathrm{T}}}
\newcommand\eqFP{\mathcal{E}^{\scriptscriptstyle \text{FP}}}
\newcommand\eqWTDiff{\mathcal{E}^{\scriptscriptstyle \text{WTDiff}}}
\newcommand\ddh{(\partial^2\!\cdot\!h)}

\makeindex

\begin{document}

\title{Nonexistence of a parent theory for general relativity and unimodular gravity}

\author{Gerardo Garc\'ia-Moreno}
\email{ggarcia@iaa.es}
\affiliation{Instituto de Astrof\'{\i}sica de Andaluc\'{\i}a (IAA-CSIC), Glorieta de la Astronom\'{\i}a, 18008 Granada, Spain}
\author{Alejandro Jim\'enez Cano}
\email{alejimcan@gmail.com}
\affiliation{Laboratory of Theoretical Physics, Institute of Physics, University of Tartu, W. Ostwaldi 1, 50411 Tartu, Estonia}

\begin{abstract}
{General relativity (GR) and unimodular gravity (UG) provide two equivalent descriptions of gravity that differ in the nature of the cosmological constant. While GR is based on the group of diffeomorphisms that permits the cosmological constant in the action, UG is based on the subgroup of volume-preserving diffeomorphisms together with Weyl transformations that forbid the presence of the cosmological constant. However, the cosmological constant reappears in UG as an integration constant so it arises as a global degree of freedom. Since gauge symmetries are simply redundancies in our description of physical systems, a natural question is whether there exists a ``parent theory'' with the full diffeomorphisms and Weyl transformations as gauge symmetries so that it reduces to GR and UG, respectively, by performing suitable (partial) gauge fixings. We will explore this question by introducing Stueckelberg fields in both GR and UG to complete the gauge symmetries in each theory to that of the would-be parent theory. Despite the dynamical equivalence of the two theories, we find that precisely the additional global degree of freedom provided by the cosmological constant in UG obstructs the construction of the parent theory.}
\end{abstract}


\maketitle

{

\hypersetup{citecolor=black, linkcolor=black,urlcolor=black}

\tableofcontents

}

\maketitle

\section{Introduction}
\label{Sec:Introduction}

Unimodular gravity (UG) is a theory that is very similar to general relativity (GR). The linearization of these theories on top of flat spacetime leads to the propagation of a massless spin-2 field. However, they are based on different gauge groups. UG admits a formulation in which it is invariant under Weyl transformations and transverse diffeomorphisms whereas GR is based on the whole group of diffeomorphisms. The differences between the two theories were explored in~\cite{Carballo-Rubio2022} (see also~\cite{Alvarez2023}), showing that they are equivalent up to the behavior of the cosmological constant at the classical level. Whereas in GR the cosmological constant is a coupling constant, in UG it is absent at the level of the action and enters as an integration constant in the equations of motion. Consequently, in UG, it behaves as a global degree of freedom, in the sense that it does not depend on the spacetime point, see~\cite{Henneaux1989,Garay2023} for further discussion. At the quantum level, the tree-level amplitudes were shown to be equivalent in both theories~\cite{Alvarez2016,CarballoRubio2019}, whereas the analysis at the loop level is more convoluted since the theory is not renormalizable and one needs to treat the theory as an effective field theory. Although there has been a lot of work in trying to discern whether the two theories were equivalent or not at the loop level~\cite{Eichhorn2013,Eichhorn2015,deBrito2019,deBrito2020,Herrero-Valea2020}, it was shown that there exists a quantization scheme in which one could show the equivalence of both theories~\cite{Percacci2017,deBrito2021}. Different choices of renormalization and quantization schemes may lead to different predictions (notice that some of the computations involve running of coupling constants, that are not directly related to observables) but the existence of a quantization scheme that provides the same predictions for both theories is enough to ensure the equivalence at the perturbative quantum level.  

Given that the two theories are so similar except for the gauge symmetries, a legitimate question is whether there exists a ``parent theory'' that is invariant under both Weyl transformations and diffeomorphisms, and such that suitable gauge fixings of the parent theory lead to UG and GR. We address such a question by introducing Stueckelberg fields in GR to make it Weyl invariant and in UG to make it invariant under longitudinal diffeomorphisms. We find that the resulting theories are not equivalent and we trace the obstruction to find such a parent theory to the presence of the global degree of freedom in UG, a global aspect that was missed in the analysis presented in~\cite{Gielen2018}.\footnote{We thank Roberto Percacci for pointing out this reference to us.}

Here is a brief outline of the article. In Sec.~\ref{Sec:Linear_Theory} we present the analysis for WTDiff and Fierz-Pauli, the linear versions of UG and GR. Subsections~\ref{Subsec:Poincaré} and~\ref{Subsec:Enlarging} provide a short introduction to the origin of gauge symmetries as a need to describe massless integer spin fields in a manifestly Lorentz invariant way. Subsection~\ref{Subsec:GlobalDof} illustrates how WTDiff differs from Fierz-Pauli in the sense that it contains an additional global degree of freedom, the linear version of the cosmological constant. Subsection~\ref{Subsec:Linear_Stueck} looks for theories that are invariant under Weyl transformations and diffeomorphisms, introducing Stueckelberg fields in WTDiff and Fierz-Pauli. The theories obtained in this way are inequivalent, and we demonstrate that it is not possible to reach a parent theory such that WTDiff and Fierz-Pauli are suitable gauge fixings of it. Section~\ref{Sec:Nonlinear} is devoted to repeating the same analysis for the nonlinear theories. Subsection~\ref{Subsec:GR_UG} contains a brief presentation of GR and UG. After that, in Subsec.~\ref{Subsec:GR_Weyl} we introduce a Stueckelberg field to make GR invariant under Weyl transformations, and in Subsec.~\ref{Subsec:UG_LDiffs} we introduce Stueckelberg fields to make UG invariant under the full set of diffeomorphisms. In Sec.~\ref{Sec:Trinity} we present the trinity formulation of UG, focusing on the metric teleparallel equivalent to UG in Subsec.~\ref{Subsec:Metric_Teleparallel_UG} and the symmetric teleparallel to UG in Subsec.~\ref{Subsec:Symmetric_Teleparallel_UG}. We show that they are inequivalent to their GR version, in the same sense that UG is inequivalent to GR. Finally, we finish in Sec.~\ref{Sec:Conclusions} by summarizing the conclusions that can be drawn up from our work.

 \paragraph*{\textbf{Notation and conventions.}}
In this article, we use the signature $(-,+,...,+)$  for the spacetime metric, and we work in natural units $c=\hbar=1$.  We also introduce $H_{(ab)}:= \frac{1}{2!}(H_{ab}+H_{ba})$ and $H_{[ab]}:= \frac{1}{2!}(H_{ab}-H_{ba})$, and similarly for an object with $n$ indices instead of 2. Latin indices ($a,b,c,d...$) refer to arbitrary coordinates in spacetime and run from $0$ to $\dimM-1$, where $\dimM$ is the dimension of the spacetime manifold that we assume to be $\dimM>3$. For the curvature tensors we use the conventions in the book of Wald~\cite{Wald1984}, {\it i.e.}, $[\nabla_a, \nabla_b] V^c =: - R_{abd}{}^c V^d$, $R_{ab}:=R_{acb}{}^c$.

\section{Linear theory of massless spin-2 particles}
\label{Sec:Linear_Theory}

\subsection{The origin of gauge symmetry}
\label{Subsec:Poincaré}

Let us begin with the spin-1 field propagating on top of flat spacetime as an illustrative example. If we have a vector field $A^{a}$, its $\dimM$ components may lead to the propagation of $\dimM$ degrees of freedom. However, from the group theory perspective, we know that if we aim to describe the propagation of massless particles, we need to ensure that it propagates only $\dimM-2$ degrees of freedom~\cite{Wigner1939,Weinberg1995}. Let us take a state that represents a plane wave
\begin{align}
    & A_{a} (x) = \epsilon_{a} (p) {\ee}^{\ii  p \cdot x}, \\
    & p^2 = 0,
\end{align}
with $p \cdot x := p_a x^a$. First of all, notice the $A^{a}$ field decomposes into $1 \oplus (\dimM-1)$ as irreducible representations of the Poincaré group. Actually, the trivial representation corresponds to the scalar encoded in $A^{a}$, namely the projection in  the direction of $p^a$, i.e. $p_{a} \epsilon^{a}(p)$ or in coordinate space $\partial_{a} A^{a}(x)$. We can remove that scalar degree of freedom by simply imposing a constraint $p_{a} \epsilon^{a}(p)=0$ which, equivalently, in coordinate space is $\partial_{a} A^{a} = 0$, the so-called Lorenz condition. However, we still need to eliminate one of the remaining states that $A^{a}$ encode in order to describe a massless particle. The first thing to notice is that there are no more Lorentz invariant constraints that we can impose on $A^{a}$ in order to remove degrees of freedom. If we intend to preserve Lorentz invariance explicitly, we cannot impose a non-Lorentz invariant constraint such as the Coulomb gauge for example. It is this absence of additional potential Lorentz-invariant constraints that leads to the introduction of gauge symmetry. Notice that if we impose the constraint $\epsilon \cdot p = 0$, any shift on the polarization vector of the form
\begin{equation}
    \epsilon_{a}  \rightarrow \epsilon'_{a}  = \epsilon_{a} + \alpha (p) p_{a}\label{eq:transfeps}
\end{equation}
will lead to a new vector that still verifies the constraint
\begin{equation}
     \epsilon \cdot p = 0  \rightarrow \epsilon \cdot p = 0, \qquad p^2  = 0.
\end{equation}
Thus, it is natural to impose that configurations related by a transformation of the form \eqref{eq:transfeps}, are physically equivalent (i.e., they belong to the same gauge orbit). We can write down this equivalence in coordinate space and it leads to the standard form of gauge symmetry:
\begin{align}
    & A_{a} \rightarrow A_{a} + \partial_{a} \alpha (x) , \nonumber \\
    & \Box \alpha = 0.
    \label{Eq:Gauge_Spin1}
\end{align}
Up to this point, we have seen that in order to describe a massless spin-1 particle through a vector field $A_{a}$, we need to ensure that it contains a dispersion relation of the form $p^2 = 0$, that the vector field is divergenceless $\partial_{a} A^{a} = 0$,  and that it displays the gauge symmetry in Eq.~\eqref{Eq:Gauge_Spin1} (which kills the additional degrees of freedom in $A^{a}$).

Let us now repeat the analysis with a tensor $h^{ab}$ aiming to describe a massless spin-2 particle. The first thing that we notice is that the tensor $h^{ab}$ contains $ \frac{1}{2} \dimM(\dimM+1) $ potential degrees of freedom, but we want it to propagate only the $ \frac{1}{2} \dimM \left( \dimM -3 \right)$~\cite{Wigner1939,Weinberg1995} associated with a massless spin-2 particle. Again, take that $h_{ab}$ represents a plane wave: 
\begin{align}
    & h_{ab} = \epsilon_{ab} (p) {\ee}^{\ii  p \cdot x}, \\
    & p^2 = 0. 
\end{align}
The first thing to notice is that again $h_{ab}$ decomposes nontrivially into irreducible representations of the Poincaré group. For instance, it contains a symmetric traceless representation, a vector representation and two scalars: i.e., the decomposition is $ \frac{1}{2} \dimM(\dimM+1) = 1 \oplus 1 \oplus (\dimM-1) \oplus \frac{1}{2}(\dimM+1)(\dimM-2)$. To be more explicit, the first scalar encoded in $h_{ab}$ is, of course, the trace $h = h^{ab} \eta_{ab}$; then we have the $\dimM$-vector $A^{a} = \partial_{b} h^{ba}$, which can be broken into the scalar $ \partial_{a} A^{a} = \partial_{a}\partial_{b} h^{ab} =:\ddh$ and the divergenceless vector, as in the previous section. In addition, we have the remaining traceless symmetric tensor $h_{ab}$ that contains more components than the ones a massless particle contains. We can remove the two scalars and the vector component by imposing suitable conditions. These conditions are that the tensor is traceless $h = 0$, which removes one of the scalars; and the Lorentz transversality condition $p_{a} \epsilon^{ab} = 0$, which removes simultaneously the other scalar $\ddh = 0$, and the vector. This is the best that we can do by using constraints that preserve Lorentz invariance. To remove the remaining components of $h_{ab}$ in a manifestly Lorentz invariant way, we need to introduce gauge symmetries again. Given that we are imposing the constraints $\epsilon^{ab} \eta_{ab} = 0$ and $p_{a} \epsilon^{ab} = 0$, any transformation on the tensor $\epsilon_{ab}$ of the form
\begin{equation}
    \epsilon_{ab} \rightarrow \epsilon'_{ab} = \epsilon_{ab} + \xi_{a} (p) p_{b} + \xi_{b} (p) p_{a}, \qquad \alpha \cdot p = 0
\end{equation}
will preserve the constraints
\begin{equation}
    p^{a} \epsilon_{ab} = \epsilon_{ab}\eta^{ab} = 0 \rightarrow p^{a} \epsilon^\prime_{ab} = \epsilon^\prime_{ab}\eta^{ab} = 0, \qquad p^2 = 0.
\end{equation}

Hence, it is natural to impose that configurations related by a transformation of this form are physically equivalent. In coordinate space, these transformations take the form:
\begin{align}
    & h_{ab} \rightarrow h_{ab} + \partial_{a} \xi_{b} + \partial_{b} \xi_{a}, \nonumber \\
    & \partial_{a} \xi^{a} =  0,\quad \Box \xi^{a}=0.  
    \label{Eq:Gauge_Spin2}
\end{align}
Thus, for describing a massless spin-2 particle through a tensor field $h_{ab}$ in a minimal way, we need to ensure that it contains a dispersion relation of the form $p^2 = 0$, the tensor needs to be traceless $h = 0$ and divergenceless $\partial_{a} h^{ab} = 0$, and realizes the gauge symmetry as in Eq.~\eqref{Eq:Gauge_Spin2}. 

\subsection{Nonminimal realizations: Enlarging the gauge symmetries}
\label{Subsec:Enlarging}

In practice, it is cumbersome to work with constrained fields (traceless, transversal, etc.). Thus, for practical purposes, working with the previous minimal constructions is not useful. The idea to avoid this is that we can eliminate the constraints that we are imposing at the expense of enlarging the gauge symmetry.

Let us illustrate this with the spin-1 field. Here there is only one way of enlarging the gauge symmetry in such a way that we only have the $A^{a}$ field as a configuration variable and we still propagate only the $\dimM-2$ desired degrees of freedom associated with the massless particle. Once we relax the condition $\partial_{a} A^{a} = 0$, we can consider more general transformations, not only those that preserve the constraint. This means that we no longer need to impose any constraint on $\alpha$, the gauge parameter of the transformation, and we can perform arbitrary transformations of the form
\begin{align}
    A_{a} \rightarrow A_{a} + \partial_{a} \alpha.
\end{align}
Now, it is easy to write down a Lagrangian for such a theory. We need to write down only the most general quadratic Lagrangian displaying this gauge symmetry and such that it gives rise to a massless dispersion relation. This is, of course, the Maxwell Lagrangian
\begin{equation}
    \mathcal{L}_{\text{Maxwell}} = - \frac{1}{4} F_{ab} F^{ab}\,,
\end{equation}
where $F_{ab} := \partial_{a} A_{b} - \partial_{b} A_{a}$.

The situation is different for the massless spin-2 particle. Here, there are two ways in which one can enlarge the amount of gauge symmetry until one works with an unconstrained field $h_{ab}$. However, it is instructive to enlarge the gauge symmetry in two steps. First of all, we relax the constraint $\partial_{a} h^{ab} = 0$. With this, we notice that we can now do more general gauge transformations, since we can relax the condition that the vector generating them obeys a wave equation, as in the previous case. The result is a transformations of the form
\begin{align}
    h_{ab} \to  h'_{ab}=h_{ab} + \partial_{a} \xi^\sT_{b} + \partial^\sT_{b} \xi_{a}, 
\end{align}
where $\xi^\sT_{a}$ is an arbitrary transverse vector field $\partial_{a} \xi^\sT_{b} \eta^{ab} = 0$, in order to preserve the traceless condition on $h_{ab}$. It is possible to write a Lagrangian that gives rise to a linear dispersion relation and implements this symmetry as long as we keep the constraint that the tensor is traceless: 
\begin{align}
    \mathcal{L}_{\text{TDiff}} = - \frac{1}{2} \partial_{a} h_{cd} \partial^{a} h^{cd} + \partial_{a} h^{ac} \partial_{b} h^{b}{}_{c}. 
\end{align}
Hereinafter we drop boundary terms depending on gauge parameters. Finally, we come to the point of relaxing the constraint $h = 0$. This can be achieved by two different enlargements of the gauge transformations. First of all, the most natural thing to do is to relax the constraint that the vectors generating the transformations are divergenceless $\partial_{a} \xi^{\sT a} = 0$, since we do not need to ensure that the trace of $h$ is preserved anymore. To put it explicitly, this means that we work with a tensor field $h_{ab}$ with the following gauge symmetry
\begin{align}
    h_{ab} \rightarrow h'_{ab} + \partial_{a} \xi_{b} + \partial_{b} \xi_{a}
\end{align}
for arbitrary $\xi_{a}$. These transformations enlarge the set of linearly realized transverse diffeomorphisms to the whole set of linearly realized diffeomorphisms. The most general Lagrangian that one can write down displaying this gauge symmetry and giving rise to a quadratic dispersion relation is the Fierz-Pauli Lagrangian~\cite{Fierz1939}
\begin{align}
    \mathcal{L}_{\text{FP}} = & - \frac{1}{2} \partial_{c} h_{ab} \partial^{c} h^{ab} + \partial_{a} h^{a}{}_{b} \partial_{c} h ^{cb} \nonumber \\
    & - \partial_{a} h^{ab} \partial_{b} h + \frac{1}{2} \partial_{a} h \partial^{a}h.
    \label{Eq:FP_Lagrangian}
\end{align}
There is though, another way of relaxing the constraint $h = 0$. We can keep the transverse diffeomorphisms and introduce the linearly realized Weyl transformations on the tensor $h_{ab}$. Explicitly, this means that we consider the following gauge group
\begin{align}
    h_{ab} \rightarrow h'_{ab} = h_{ab}+ \partial_{a} \xi^\sT_{b} + \partial_{b} \xi^\sT_{a} + \chi \eta_{ab}, 
\end{align}
where the $ \chi$ is an arbitrary scalar function and $\xi^\sT_{a}$ is transverse (divergenceless). Imposing that the set of linear Weyl transformations and transverse diffeomorphisms are realized at the Lagrangian level together with the massless condition, leads up to a global constant to the so-called WTDiff Lagrangian~\cite{Alvarez2006}
\begin{align}
    \mathcal{L}_{\text{WTDiff}} = & - \frac{1}{2} \partial_{c} h_{ab} \partial^{c} h^{ab} + \partial_{a} h^{a}{}_{b} \partial_{c} h^{cb} \nonumber \\
    & - \frac{2}{\dimM} \partial_{a}h^{ab} \partial_{b} h + \frac{\dimM+2}{2\dimM^2} \partial_{a} h \partial^{a} h .
    \label{Eq:WTDiff_Lagrangian}
\end{align}

\subsection{Same local degrees of freedom, different number of global degrees of freedom}
\label{Subsec:GlobalDof}
At this point, one may lead to the conclusion that both theories need to be equivalent since they lead to the propagation of the same number of local degrees of freedom. However, as we will now discuss, there is a mismatch in the number of global degrees of freedom. 
\paragraph*{\textbf{Fierz-Pauli case.}}
To see this, let us begin with the Fierz-Pauli theory. The corresponding equations of motion are $\eqFP_{ab} = 0$, where $\eqFP_{ab}$ is given by
\begin{align}
    \eqFP_{ab} &:=  \Box h_{ab} - 2 \partial_{(a | } \partial_{c} h^{c}{}_{| b)}  \nonumber \\ 
    & \quad + \partial_{a} \partial_{b} h + \eta_{ab} \ddh - \eta_{ab} \Box h ,
    \label{Eq:Fierz_Pauli_Eqs}
\end{align}
which fulfills
\begin{align}
    \eta^{ab}\eqFP_{ab} = & (\dimM-2)\Big[\ddh-\Box h\Big] \\
    \partial^b\eqFP_{ab}= & 0.
\end{align}
We can now perform a gauge fixing to reach $h = 0$ and $\partial_{a} h^{ab} = 0$. Notice that given a value for $\partial_{a} h^{ab}$ and $ h$, we can always perform a diffeomorphism:
\begin{align}
    \partial_{a} h^{\prime ab} &= \partial_{a} h^{ab} + \Box \xi^{b} + \partial^{b} \partial_{a} \xi^{a}\,, \\
    h' &= h + 2 \partial_{a} \xi^{a}\,,
\end{align}
such that $\partial_{a} h^{\prime ab} = h' = 0$. This leads to the following system of equations for $\xi^a$ that always admits a solution:
\begin{align}
    & \partial_{a} \xi^{a} = -\frac{1}{2} h, \\
    & \Box \xi^{a} = \frac{1}{2} \partial^{a} h - \partial_{b} h^{ba} .
\end{align}
Hence, we can always make a gauge fixing such that Eqs.~\eqref{Eq:Fierz_Pauli_Eqs} reduce to a sourceless wave equation for a transverse traceless tensor:
\begin{align}
    & \Box h_{ab} = 0, \\
    & \partial_{a} h^{ab} = h = 0. 
\end{align}

\paragraph*{\textbf{WTDiff case.}}
Following~\cite{Bonifacio2015}, we can try to do the same for the WTDiff theory \eqref{Eq:WTDiff_Lagrangian}. In this case, the dynamical equations are $\eqWTDiff_{ab}  =0$\footnote{
    Because of the Weyl symmetry, this equation is actually independent of the trace of $h_{ab}$. Indeed, if we decompose $h_{ab}=\hat{h}_{ab}+\frac{1}{\dimM}\eta_{ab}h$ (where by construction $\eta^{ab}\hat{h}_{ab}=0$), we find the identity
    \[
        \eqWTDiff_{ab} = \Box \hat{h}_{ab} - 2 \partial_{(a|} \partial_{c} \hat{h}^{c}{}_{|b)}  + \frac{2}{\dimM} \eta_{ab} (\partial^2\!\cdot\!\hat{h}) \,.
    \]
}, with
\begin{align}
    \eqWTDiff_{ab}& :=  \Box h_{ab} - 2 \partial_{(a|} \partial_{c} h^{c}{}_{|b)} + \frac{2}{\dimM} \partial_{a} \partial_{b} h \nonumber \\
    &\quad + \frac{2}{\dimM} \eta_{ab} \ddh - \frac{\dimM+2}{\dimM^2} \eta_{ab} \Box h .
    \label{Eq:WTDiff_Eqs}
\end{align}
which fulfills
\begin{align}
    \eta^{ab}\eqWTDiff_{ab} & = 0 \\
    \partial^b\eqWTDiff_{ab} & = -\frac{\dimM-2} {\dimM}\partial_a\Big[\ddh-\frac{1}{\dimM}\Box h\Big]\,.
\end{align}
From the second one we can derive the general result that, on-shell, the quantity in the square bracket must be constant, i.e.,
\begin{equation}
    \ddh-\frac{1}{\dimM}\Box h = c\,,\label{eq:conddivWTDiff}
\end{equation}
where $c$ is the integration constant. 

Now one can try to find a gauge leading to $\Box h_{ab} = J_{ab}$ such that the field $h_{ab}$ is traceless and transverse and such that the source $J_{ab}$ is independent of $h_{ab}$. However, it is not difficult to check that due to the presence of $c$ one cannot achieve both conditions simultaneously. Let us show this in more detail.

First, we perform a Weyl gauge transformation to reach $h = 0$. It is always possible to do so since under a Weyl transformation, we have
\begin{align}
    h' = h + \dimM \chi, 
\end{align}
and taking $\chi = - h/{\dimM}$, we reach $h'=0$. The condition \eqref{eq:conddivWTDiff} implies then that, in this gauge, $\ddh = c$. This can be integrated to obtain an expression for the divergence of $h^{ab}$
\begin{align}
    \partial_{a} h^{ab} = \frac{c}{\dimM} x^{b} + a^{\sT b}, 
\end{align}
where $a^\sT_{a}$ is any arbitrary divergenceless vector field $\partial^{a} a^\sT_{a} = 0$. The vector $a^\sT_{a}$ can be removed it through a gauge transformation. To see this, notice that under a transversal diffeomorphism,
\begin{align}
    \partial_{a} h^{\prime ab} = \partial_{a} h^{ab} + \Box \xi^{\sT b}, 
\end{align}
so the condition $\partial_{a} h^{\prime ab} = \frac{c}{\dimM} x^{b}$ leads to an equation for $\xi^{\sT b}$ that always admits a solution:
\begin{align}
    \Box \xi^{\sT}_{b} = - a^\sT_{b}. 
\end{align}
Hence, in this gauge, we end up again with a sourceless wave equation for $h_{ab}$, 
\begin{align}
    \Box h_{ab} = 0,\label{eq:boxhfixedWTDiff}
\end{align}
with
\begin{align}
    \partial_{a} h^{ab} = \frac{c}{\dimM} x^{b}\,,\qquad\qquad h=0\,.
\end{align}
Note that the transversality condition is violated for nonvanishing $c$.

To explore the other possibility, we can make a field redefinition
\begin{align}
    h_{ab} \mapsto H_{ab} + \frac{1}{2} \frac{c}{\dimM} \eta_{ab} x^2,
\end{align}
in such a way that this new tensor $H_{ab}$ is transverse, although it is not traceless anymore:
\begin{equation}
    \partial_{a} H^{ab} = 0, \qquad  H_c{}^c = - \frac{1}{2} c x^2. 
\end{equation}
The equation of motion \eqref{eq:boxhfixedWTDiff} in terms of $H_{ab}$ contains a constant source term proportional to $c$:
\begin{align}
    \Box H_{ab} = - c \eta_{ab}\,.
    \label{Eq:Dhceta}
\end{align}

It is also remarkable that Eq. \eqref{Eq:Dhceta} coincides with the lowest-order contribution to the graviton equation in the presence of a cosmological constant $c$ (obtained, for instance, from the linearization of GR). We will see later that in the full nonlinear theory, the integration constant $c$ is promoted to be the full cosmological constant entering Einstein equations. Since this $c$ is arbitrary, it corresponds to an additional degree of freedom of WTDiff, which needs to supplement the set of initial conditions.

Up to this point, we have noticed that although both theories lead to the propagation of the same number of local degrees of freedom, they are not equivalent because there is one additional number required to specify the initial conditions for WTDiff, $c$. In that sense, they cannot be regarded as two formulations of the same theory. Let us further dig into this question in the following subsection.

\subsection{Stueckelberg-ing}
\label{Subsec:Linear_Stueck}

It is always possible to enlarge the set of gauge symmetries of a theory at the expense of introducing additional fields that are often called Stueckelberg fields, being the opposite to minimal formulations in terms of the number of fields. One natural question that we can ask is whether there exists a ``parent'' theory that involves more fields, not only the tensor $h_{ab}$, but is invariant both under general diffeomorphisms and Weyl transformations acting linearly on the tensor $h_{ab}$ such that Fierz-Pauli and WTDiff are suitable gauge fixings of both of them. Although based on our discussion about the global mode of WTDiff we may anticipate that this is not possible, it is interesting to perform the analysis explicitly. 

\paragraph*{\textbf{Fierz-Pauli case.}}
Let us begin with Fierz-Pauli. The parent theory that we seek would be a theory in which the action is invariant under linearized Weyl transformations on the tensor $h_{ab}$ in addition to linearized diffeomorphisms. The way to achieve this through a Stueckelberg field is to introduce a field $\varphi$ in the action by making the replacement $h_{ab} \rightarrow h_{ab} + \varphi \eta_{ab}$. In this way, we realize the Weyl symmetry in the $\varphi$ field as a shift symmetry
\begin{align}
    h_{ab} &\to h'_{ab} = h_{ab} + \chi \eta_{ab}\,, \\
     \varphi &\to \varphi'=\varphi - \chi\,. 
\end{align}
After introducing the field $\varphi$, we are led to the following Lagrangian: 
\begin{align}
     \mathcal{L}_{\text{FP-St}} &= \mathcal{L}_{\text{FP}} -  (\dimM - 2) \partial_{a}h^{ab} \partial_{b} \varphi + (\dimM - 2) \partial_{a}h \partial^{a} \varphi\nonumber \\ 
     & \quad  + \frac{1}{2} (\dimM - 1)(\dimM - 2) \partial_{a} \varphi \partial^{a} \varphi\,, 
    \label{Eq:Stueckelberg_GR}
\end{align}
whose equations of motion are
\begin{align}
    \Box \varphi - \frac{1}{(\dimM - 1)(\dimM-2)} \eta^{ab}\eqFP_{ab} = 0\,, \\
   \eqFP_{ab} + (\dimM-2)( \partial_{a} \partial_{b} \varphi - \eta_{ab}\Box \varphi )= 0\,. \label{eq:StFP}
\end{align}
One important thing to notice is that the equation of $\varphi$ is the trace of the second one, so it is redundant (in any gauge). Therefore, from now on we just keep Eq.~\eqref{eq:StFP}.

It is immediate to see that the unitary gauge $\varphi=0$ is a legal one, since the equation of motion of $\varphi$ becomes redundant. Here by legal we mean that fixing the gauge in the action and fixing it in the equations lead to the same dynamics. However we want to emphasize that what we call here illegal gauge fixings can perfectly be used at the level of the equations of motion (as long as there exists a gauge transformation leading to them) without any inconsistency. It is only incorrect to use them at the level of the action, in the sense of imposing the conditions directly on the fields before performing the variation that leads to the equations of motion of the theory. Notice that by ``gauge fixing at the level of the action" we do not mean to introduce a Lagrange multiplier in the Lagrangian to enforce the constraint. Such a procedure would introduce additional (algebraic) equations of motions that would follow from the variation of the Lagrange multiplier fields and would lead to the same dynamics reached by fixing the gauge directly in the equations of motion. We give a more elaborate analysis of legal and illegal gauge fixings in Appendix~\ref{Sec:Illegal_gauge}.

The question now is whether one can find a legal gauge that leads to the WTDiff dynamics. 
In principle, fixing the gauge at the level of the action to be $\varphi = - \frac{1}{\dimM} h $ leads to the WTDiff Lagrangian  \eqref{Eq:WTDiff_Lagrangian}. However, this is an illegal gauge fixing, as we will see below.  To show that it is always possible to reach this gauge, we recall that under a generic gauge transformation the fields change as
\begin{align}
    h_{ab}&\to h'_{ab} = h_{ab} + 2 \partial_{(a} \xi_{b)} + \chi \eta_{ab} \noindent\\
    \varphi&\to \varphi' = \varphi - \chi \,.\label{eq:totalsym-FPS}
\end{align}
We want to see if for generic $\{\varphi, h_{ab}\}$ we can find $\{\xi^{a}, \chi\}$ such that we end up with $ \varphi' = - h'/\dimM$. The latter condition fixes the divergence of the vector $\xi_{a}$:
\begin{align}
    \partial_{a} \xi^{a} = - \frac{1}{2} h - \frac{\dimM}{2} \varphi.
\end{align}
A trivial example of such a vector field would be
\begin{align}
    \xi^{a} =  \delta^{a}{}_0 \int^{t}_{t_0} \dd t'  \left( - \frac{1}{2} h (t',x^i) - \frac{\dimM}{2} \varphi(t',x^i) \right).
    \label{Eq:Diff_Transf}
\end{align}
Let us now show that this gauge fixing is illegal. In this gauge, the only independent equation of motion \eqref{eq:StFP} reduces to
\begin{equation}
    \eqWTDiff_{ab}+\frac{\dimM-2}{\dimM}\eta_{ab}\big[ \ddh - \frac{1}{\dimM} \Box h \big]=0\,. \label{eq:StFPFix}
\end{equation}
Since $\eqWTDiff_{ab}$ is traceless and the second term in \eqref{eq:StFPFix} is pure trace, this equation is fulfilled if and only if both terms in \eqref{eq:StFPFix} vanish independently. The first one leads to the WTDiff equations and this implies \eqref{eq:conddivWTDiff} with the appearance of the integration constant $c$. However, the second term in \eqref{eq:StFPFix} enforces $c=0$. This proves that the gauge fixing $\varphi = - h/n$ is illegal since the equations resulting from fixing the gauge at the level of equations do not reproduce the dynamics of the WTDiff action, which admits an arbitrary $c$.

\paragraph*{\textbf{WTDiff case.}}
We can repeat the same analysis for WTDiff. We can introduce a Stueckelberg field to enlarge the gauge symmetries from Weyl and transverse diffeomorphisms to Weyl and general diffeomorphisms, at the expense of introducing an additional Stueckelberg field. We want to distinguish between longitudinal and transverse diffeomorphisms. For such a purpose, we notice that for an arbitrary diffeomorphism, we can always decompose the vector $\xi_{a}$ into a transverse part and a longitudinal part:
\begin{equation}
    \xi_{a} = \xi_{a}^\sT + \partial_{a} \sigma \qquad (\partial^{a} \xi_{a}^\sT = 0)\,.
\end{equation}
When we expand such a transformation acting on $h_{ab}$ we have
\begin{align}
    h_{ab} \rightarrow h'_{ab} = h_{ab} + \partial_{a} \xi^\sT_{b} + \partial_{b} \xi^\sT_{a} + 2 \partial_{a} \partial_{b} \sigma.\label{eq:htransfxi}
\end{align}
To make WTDiff invariant also under longitudinal diffeomorphisms~\eqref{Eq:WTDiff_Lagrangian}, we need to introduce a Stueckelberg field making the replacement
\begin{align}
    h_{ab} \to h_{ab} + 2 \partial_{a} \partial_{b} \varphi. \label{eq:stuckWTDiff}
\end{align}
The resulting Lagrangian, after some integrations by parts, is
\begin{align}
    \mathcal{L}_\text{WTDiff-St}   & =   \mathcal{L}_\text{WTDiff}+ 2 \frac{(\dimM-2)}{\dimM} \partial_{a} h^{ab} \partial_{b} \Box \varphi \nonumber \\
     - 2 \frac{\dimM-2}{\dimM^2} & \Box \partial^{a} \varphi \partial_{a} h + 2 \frac{(\dimM-1)(\dimM-2)}{\dimM^2} \Box \partial_{a} \varphi  \Box \partial^{a} \Box \varphi\,,
    \label{Eq:WTDiff-St_lagrangian}
\end{align}
whose equations are
\begin{align}
    \partial^a\partial^b\eqWTDiff_{ab} - \frac{2(\dimM-1)(\dimM-2)}{\dimM^2}\Box^3\varphi=0\,, \\
   \eqWTDiff_{ab} - \frac{2(\dimM-2)}{\dimM}\left[  \partial_{a}\partial_{b} \Box\varphi -\frac{1}{\dimM}\eta_{ab}\Box^2\varphi\right] = 0\,.\label{eq:StWTDiff}
\end{align}
Observe that, again, the equation of $\varphi$ can be obtained by taking $\partial^a\partial^b$ in the second one so it is redundant (for any gauge), so we can just keep the second one \eqref{eq:StWTDiff}.

As a consequence of the substitution \eqref{eq:stuckWTDiff}, the simultaneous transformation \eqref{eq:htransfxi} and
\begin{align}
    \varphi  \to \varphi' = \varphi - \sigma,
\end{align}
becomes a symmetry of the theory. Therefore, Eq.~\eqref{Eq:WTDiff-St_lagrangian} exhibits the following symmetry:
\begin{align}
    h_{ab}&\to h'_{ab} = h_{ab} +\partial_{a} \xi^\sT_{b} + \partial_{b} \xi^\sT_{a} + 2 \partial_{a} \partial_{b} \sigma + \chi \eta_{ab} \noindent\\
    \varphi&\to \varphi' = \varphi - \sigma \,.\label{eq:totalsym-WTDiffS}
\end{align}
Notice that this is not the same as the one found in the Fierz-Pauli case, Eq.~\eqref{eq:totalsym-FPS}. This shows explicitly two different ways of realizing the gauge group of Weyl transformations and full diffeomorphisms using only a scalar field and a tensor $h_{ab}$. One could wonder whether there is some kind of field redefinition that maps one theory into the other one, but we will now argue why this is not possible because there is a mismatch in the global degrees of freedom of the theory.

Let us now show that it is not possible to find a legal gauge fixing of the Weyl symmetry leading to the Fierz-Pauli theory. The gauge in which $\Box \varphi = - \frac{1}{2} h$ performed at the level of the action leads to the Fierz-Pauli Lagrangian. This means that performing such gauge fixing we would lose the global degree of freedom associated with the constant $c$. This seems contradictory, so let us analyze in detail. Introducing this into the action~\eqref{Eq:WTDiff-St_lagrangian}, one recovers the Fierz-Pauli Lagrangian from Eq.~\eqref{Eq:FP_Lagrangian}. First of all, it is always possible to reach this gauge through a suitable Weyl transformation: under a Weyl transformation, the trace $h$ and $\varphi$ change as
\begin{align}
    & h \rightarrow h' = h +  \dimM \chi , \\
    & \varphi \rightarrow \varphi' = \varphi ,
\end{align}
where $\chi$ is the gauge parameter of the transformation. By imposing $\Box \varphi' = - \frac{1}{2} h'$, we find the $\chi$ that generates the gauge transformation that we are looking for is
\begin{align}
    \chi = - \frac{1}{\dimM} \left( h + 2 \Box \varphi \right). 
\end{align} 
Thus, it is always possible to reach that gauge. To see that this is illegal, we realize that the only independent equation \eqref{eq:StWTDiff} after evaluating the gauge $\Box \varphi = - \frac{1}{2} h$ becomes
\begin{align}
    &\Box \big(\ddh - \Box h \big)=0\,, \label{eq:FPx1}\\
    &\eqFP_{ab} - \frac{\dimM-2}{\dimM} \eta_{ab}\big(\ddh -\Box h \big) = 0 \,.
\end{align}
By taking the divergence of the second one, and recalling that the $\eqFP_{ab}$ part is divergenceless, we find that the combination $\big(\ddh -\Box h \big)$ must be constant. Therefore, Eq.~\eqref{eq:FPx1} is redundant and the system of equations becomes:
\begin{equation}
    \eqFP_{ab} =  c\eta_{ab} \,,
    \label{Eq:fixing_constant}
\end{equation}
for some constant $c$.

The equations that follow from the Lagrangian after gauge fixing are not the equations that we obtain by performing the gauge fixing directly at the level of the equations of motion. To be more precise, since the gauge-fixed Lagrangian is Fierz-Pauli, the equations that follow from that Lagrangian do not contain an integration constant. However, the equations obtained by performing the gauge fixing at the level of the equations of motion do contain such a constant~\eqref{Eq:fixing_constant}. Thus, the gauge fixing is illegal in the sense that we explain in Appendix~\ref{Sec:Illegal_gauge}.

At the linear level, which we have analyzed until now, we conclude that there is not a parent theory as we have dubbed it, from which Fierz-Pauli and WTDiff are two suitable (legal) gauge fixings. The reason behind it is that while the two theories agree on the local degrees of freedom that they propagate, WTDiff contains an extra global degree of freedom that can be understood as the linear version of the cosmological constant.

One might wonder whether this mismatch arises from the fact that we are comparing WTDiff with Fierz-Pauli without a cosmological constant and we should add it in our analysis. Another possibility is that we are always focusing on the trivial flat spacetime background, which is only acceptable in WTDiff as long as the initial conditions fix $c = 0$, and we need to consider the theories in arbitrary backgrounds. The reason for the impossibility to find a parent theory is actually that UG contains all the possible values of the cosmological constant within a single theory, whereas in Fierz-Pauli different values of the cosmological constant correspond to different theories. To remove any doubt, we extend the analysis considering the nonlinear versions of these linearized theories in the following sections and we find the same result.

\section{Nonlinear theories}
\label{Sec:Nonlinear}

GR and UG are the nonlinear completions of Fierz-Pauli and the linearized theory of WTDiff. Both of them propagate two local degrees of freedom, just as their linear versions. However, the difference due to the mismatch in the global degrees of freedom is still present. UG contains a fiduciary nondynamical background structure (see Appendix~\ref{sec:diffs}), a nondynamical volume form, something that is tightly related to this fact. For a comprehensive review of UG and its comparison with GR, see~\cite{Carballo-Rubio2022}.

Because of this global degree of freedom and based on the analysis at the linear level, we expect that it is not possible to introduce Stueckelberg fields for GR and UG, enlarging the gauge symmetry by adding Weyl transformations and longitudinal diffeomorphisms, in such a way that we go to the same parent theory.

\subsection{General relativity and unimodular gravity}
\label{Subsec:GR_UG}

Let us quickly review general relativity and unimodular gravity in order to settle the notation that we will be using. Let us begin with GR. GR is given by the equations following from the Einstein-Hilbert action with cosmological constant
\begin{align}
    S_{\text{GR-}\Lambda}[\boldsymbol{g}] = \frac{1}{2 \kappa^2} \int \dd^{\dimM}x\, \sqrt{\abs{g}} \left( - 2 \Lambda  + R [\boldsymbol{g}] \right),
    \label{eq:actionGR}
\end{align}
and the case with vanishing cosmological constant will be represented as $S_{\text{GR}}$. The equations of motion that follow from this action are Einstein equations:
\begin{align}
    R_{ab} [\boldsymbol{g}] - \frac{1}{2} R [\boldsymbol{g}] g_{ab} + \Lambda g_{ab} = 0.  
\end{align}
The action is invariant under general coordinate transformations (GCT), and since there is no background structure, GCT are identified with active diffeomorphisms (Diff).

Unimodular gravity can be understood as the theory that is derived from the Einstein-Hilbert action principle by imposing the constraint that the determinant of the metric is fixed to be a nondynamical background volume form $\boldsymbol{\omega}:=\omega (x)~\dd^{\dimM}x$ (see Section \ref{sec:diffs}): 
\begin{align}
    S_{\text{UG-}\lambda}[\boldsymbol{g},\lambda] &=  \frac{1}{2 \kappa^2} \int \dd^{\dimM}x\, \Bigg[ \sqrt{\abs{g}} \left( - 2 \Lambda  + R [\boldsymbol{g}] \right) \nonumber \\
    & \quad + \lambda \left( \sqrt{\abs{g}} - \omega  \right) \Bigg],\label{eq:actionUGla}
\end{align}
where $\lambda(x)$ is a suitable Lagrange multiplier. The variation of this action yields to the traceless version of Einstein equations. However, it is convenient to work with unconstrained variables again, at the expense of enlarging the gauge symmetry. This can be achieved by introducing the auxiliary metric tensor $\boldsymbol{\tilde{g}}$ whose components are
\begin{align}
    \tilde{g}_{ab} = g_{ab} \left( \frac{\omega^2}{\abs{g}} \right)^{\frac{1}{\dimM}}, \label{eq:gtilde}
\end{align}
which by construction satisfies
\begin{equation}
    \sqrt{\abs{\tilde{g}}}= \omega \,.\label{eq:detgtilde}
\end{equation}
The reason for introducing this metric is that, when performing variations, only the $\boldsymbol{g}$-metric is varied, since $\omega$ is not a dynamical field. Thus, any object written in terms of the tensor $\boldsymbol{\tilde{g}}$ that we vary with respect to $\boldsymbol{g}$ will automatically produce the traceless variation with respect to the tensor $\boldsymbol{\tilde{g}}$:
\begin{align}
    \delta \tilde{g}^{ab} =\left( \frac{\omega^2}{\abs{g}} \right)^{-\frac{1}{\dimM}} \left( \delta g^{ab} - \frac{1}{\dimM} \tilde{g}^{ab} \tilde{g}_{cd} \delta g^{cd}\right).
\end{align}
In terms of this metric, the unconstrained action of UG can be expressed as
\begin{equation}
    S_{\text{UG}} [\boldsymbol{g}; \boldsymbol{\omega}] = \frac{1}{2 \kappa^2} \int \dd^{\dimM}x\, \omega R  [\boldsymbol{\tilde{g}}(\boldsymbol{g},\boldsymbol{\omega})]. \label{eq:actionUG}
\end{equation}
We separate the background volume form with a semicolon instead of a comma from the rest of the fields to indicate clearly that it is nondynamical. We will follow this notation from now on. The equations of motion that follow from this action principle are the traceless version of Einstein-equations
\begin{equation}
    R_{ab} [\boldsymbol{\tilde{g}}] - \frac{1}{\dimM} R [\boldsymbol{\tilde{g}}] \tilde{g}_{ab} = 0,
\end{equation}
which, upon using Bianchi identities, become the Einstein equations with the cosmological constant entering as an arbitrary integration constant \cite{Carballo-Rubio2022}:
\begin{equation}
    R_{ab} [\boldsymbol{\tilde{g}}] - \frac{1}{2} R [\boldsymbol{\tilde{g}}] \tilde{g}_{ab} + \Lambda \tilde{g}_{ab} = 0. \label{eq:eqUGLambda}
\end{equation}
Finally, we comment on the local symmetries of this theory. Contrary to the GR case, UG presents a background structure, $\omega(x)$, which remains unaffected under active diffeomorphisms (i.e., it transforms as \eqref{eq:bgstructransf}). With this in mind, we proceed to enumerate the symmetries of the UG action:
\begin{enumerate}
    \item General coordinate transformations ($\mathrm{GCT}$). This is clear since the theory is written in terms of two tensor quantities: the metric $g_{ab}$ and the scalar density $\omega$, in addition to the matter fields (we assume they are introduced as usual via tensor-valued quantities).

    \item Active diffeomorphisms of unit determinant  ($\mathrm{TDiff}$). These diffeomorphisms are also called transversal or volume-preserving and they are defined by \eqref{eq:activediff} together with the condition $\abs{J}=1$. To be precise, by this symmetry we mean the realization in which $\omega(x)$ is treated as a background structure in the sense of Appendix~\ref{sec:diffs}.

    \item Weyl reescalings of the metric. By this we mean transformations of the form
    \begin{align}
        g_{ab}(x) &  \to {\ee}^{\phi(x)} g_{ab}(x) \nonumber \\
        \omega(x) & \to \omega(x)\,.
    \end{align}
    This symmetry is explicitly manifest in formulation \eqref{eq:actionUG} of UG. This action depends only on the metric $\boldsymbol{g}$ through the auxiliary metric $\boldsymbol{\tilde{g}}$, which is constructed with the Weyl-invariant quantities $\omega$ and $g_{ab}/\abs{g}^{1/\dimM}$.
    
\end{enumerate}

The gauge symmetries (those besides $\mathrm{GCT}$) that this theory displays act at the infinitesimal level as follows: 
\begin{equation}
    \delta_{\xi} \tilde{g}_{ab} = \tilde{\nabla}_{a} \xi_{b} + \tilde{\nabla}_{b} \xi_{a} + \chi \tilde{g}_{ab} , \qquad \tilde{\nabla}_{a} \xi^{a} = 0, 
    \label{Eq:TDiffs}
\end{equation}
where $\tilde{\nabla}$ is the Levi-Civita connection associated with the metric $\boldsymbol{\tilde{g}}$. If we couple to the matter fields through the metric $\boldsymbol{\tilde{g}}$ as we do in GR, they automatically inherit the invariance under $\mathrm{TDiff}$ and Weyl transformations of the metric (notice that $\boldsymbol{\tilde{g}}$ is already Weyl invariant). 

\subsection{Making GR gauge invariant under Weyl rescalings of the metric}
\label{Subsec:GR_Weyl}

Let us begin with the GR action with an arbitrary cosmological constant $\Lambda$, Eq.~\eqref{eq:actionGR}. We want this action to be invariant under Weyl rescalings of the metric that are transformations that act on the metric as 
\begin{align}
    g_{ab} \rightarrow {\ee}^{2 \phi (x)} g_{ab}. 
    \label{Eq:Weyl_Resc}
\end{align}
For such a purpose, we introduce a new field $\pi(x)$ in the action to build the new action depending on both fields $\pi$ and the metric as
\begin{align}
    S_{\text{GR-} \Lambda \text{-St}}  [\boldsymbol{g},\pi] = S_{\text{GR-}\Lambda} [ {\ee}^{2 \pi} \boldsymbol{g} ]. \label{eq:GRStdef}
\end{align}
For this action to be invariant under Weyl rescalings of the metric we need that the $\pi$ field transforms through a shift
\begin{align}
    \pi \rightarrow \pi - \phi,
\end{align}
which in conjunction with Eq.~\eqref{Eq:Weyl_Resc} allows us to deduce that the combination ${\ee}^{2 \pi} g_{ab} $ is invariant. Hence, the action $S_{\text{GR-} \Lambda \text{-St}}[\boldsymbol{g},\pi]$ is Weyl invariant. We can use the transformation properties of the Ricci scalar under these rescalings of the metric to write down the action $S_{\text{GR-} \Lambda \text{-St}}[\boldsymbol{g},\pi]$ explicitly in terms of the Stueckelberg field $\pi$ and the metric $\boldsymbol{g}$. Under a transformation $\boldsymbol{g} \rightarrow \ee^{2\phi} \boldsymbol{g}$ the Ricci scalar transforms as (Eq. D.9 from~\cite{Wald1984} with  $\Omega = \ee^{\phi}$)
\begin{widetext}
\begin{align}
    R\left[ {\ee}^{2\phi} \boldsymbol{g} \right] = {\ee}^{-2\phi} \Big( R\left[ \boldsymbol{g} \right] - 2 (\dimM - 1) g^{ab}\nabla_a \nabla_b \phi - (\dimM - 1) (\dimM-2) g^{ab} \nabla_a \phi \nabla_b  \phi \Big), 
\end{align}
\end{widetext}
where $\Box := g^{ab} \nabla_a \nabla_b$ and $ (\partial \phi )^2 = g^{ab} \nabla_a \phi \nabla_b \phi$.

The change in the determinant is a simple exponential factor in $\pi$: $\abs{g} \rightarrow {\ee}^{2 \pi \dimM} \abs{g}$. We obtain the following action after integration by parts
\begin{widetext}
\begin{align}
    S_{\Lambda\text{GR-St}} [\boldsymbol{g},\pi] = \frac{1}{2 \kappa^2} \int \dd^{\dimM}x\, \sqrt{\abs{g}} {\ee}^{ (\dimM-2) \pi} \left[ - 2 \Lambda {\ee}^{2 \pi} + R + (\dimM - 1) (\dimM-2) ( \partial \pi)^2 \right]. 
    \label{Eq:Stueckelberg}
\end{align}
\end{widetext}
Here we may want to go to a gauge in which the theory is still invariant TDiff and Weyl transformations but not under longitudinal Diff. To achieve this, we would do the following. We can write down the metric as follows: 
\begin{align}
    g_{ab} = \abs{g}^{\frac{1}{\dimM}} \frac{g_{ab}}{\abs{g}^{\frac{1}{\dimM}}} = \abs{g}^{\frac{1}{\dimM}} g'_{ab},
\end{align}
where we have introduced the auxiliary field $g'_{ab} := g_{ab} / \abs{g}^{\frac{1}{\dimM}}$ which is automatically invariant under Weyl rescalings, and transforms as a tensor under TDiff, though not under longitudinal Diff. Now, to end up in an action that is only Weyl and TDiff invariant, we fix the longitudinal Diff by imposing 
\begin{align}
    {\ee}^{2 \pi } \abs{g}^{\frac{1}{\dimM}}  = \omega^{\frac{2}{\dimM}}.
    \label{eq:gaugefix}
\end{align}
where $\omega$ is an arbitrary (but fixed) background structure. Consistency of \eqref{eq:gaugefix} requires $\omega$ to be a scalar density with the same weight as $\sqrt{\abs{g}}$ under $\mathrm{GCT}$.

The question now is whether it is possible to reach this condition through a longitudinal Diff. The answer is in the affirmative, since for an arbitrary value of the factor ${\ee}^{2 \pi} \abs{g}^{\frac{1}{\dimM}}$, let us call it a density function $F(x)^{\frac{2}{\dimM}}$, a longitudinal Diff leads to the following transformation rule:
\begin{align}
    {\ee}^{2 \pi} \abs{g}^{\frac{1}{\dimM}}=F^{\frac{2}{\dimM}}\quad  \rightarrow \quad \big(F\abs{J}\big)^{\frac{2}{\dimM}}  , 
\end{align}
where we have used that ${\ee}^{2 \pi}$ is a scalar and that $\abs{g}$ picks a factor of $\abs{J}$. Thus, to reach the gauge \eqref{eq:gaugefix}, we just take: 
\begin{align}
    \abs{J}=\frac{\omega}{F}= \frac{\omega}{ \sqrt{\abs{g}}\, {\ee}^{\dimM \pi}} 
\end{align}
If we fix this gauge at the level of the action, and call 
\begin{align}
    \tilde{g}_{ab}  := \omega^{\frac{2}{\dimM}} g'_{ab}  =  \left( \frac{\abs{g}}{\omega^2} \right)^{-\frac{1}{\dimM}}g_{ab}  , 
\end{align}
we get:
\begin{align}       
    & S_{\text{GR-}\Lambda\text{-St}} [\boldsymbol{g}|_{\eqref{eq:gaugefix}},\pi]    
     = \frac{1}{2 \kappa^2} \int \dd^{\dimM}x\, \omega R  [\boldsymbol{\tilde{g}}]-\frac{\Lambda}{ \kappa^2} \int \dd^{\dimM}x\, \omega  \,, \label{eq:SLGRStfixed}
\end{align}
where in the first equality we used \eqref{eq:GRStdef} and in the second one we took into account \eqref{eq:detgtilde}. In \eqref{eq:SLGRStfixed} we see that, as a result of the gauge fixing, we obtain UG action \eqref{eq:actionUG} plus a non-dynamical constant contribution.

Fixing this gauge at the level of the action leads to UG, but this gauge fixing turns out to be illegal. To see it, we can compute the equations of motion from~\eqref{Eq:Stueckelberg}. By varying with respect to $\pi$ and $g^{ab}$, we get (after recasting the equations a bit)
\begin{align}
    0&= (\dimM-2) R [{\ee}^{2 \pi} \boldsymbol{g}] -2 \dimM \Lambda \, , \label{eq:eqpiGRS}\\
    0&= R_{ab}[\boldsymbol{g}] - \frac{1}{2} R[\boldsymbol{g}]  g_{ab} + \Lambda {\ee}^{2 \pi}g_{ab}- (\dimM-2) \nabla_{a} \nabla_{b} \pi  \nonumber \\
    &\quad +(\dimM-2) g_{ab} \Box \pi + (\dimM-2) \partial_{a} \pi \partial_{b} \pi \nonumber \\
    &\quad+ \frac{1}{2} (\dimM-2)(\dimM-3) g_{ab} (\partial \pi)^2.\label{eq:eqgGRS}
\end{align}
Notice that the first one implies  the following constraint between the Ricci scalar $R [ {\ee}^{2 \pi} \boldsymbol{g} ]$ and the cosmological constant
\begin{align}
    R [ {\ee}^{2 \pi} \boldsymbol{g} ] = \frac{2\dimM}{\dimM-2}\Lambda\, . 
\end{align}
Fixing the unitary gauge $\pi = 0$ in \eqref{eq:eqpiGRS} and \eqref{eq:eqgGRS} leads to Einstein equations with cosmological constant $\Lambda$ and its trace, meaning that both of them are not independent, as it should be. Similarly, in the gauge $ {\ee}^{\dimM \pi} = \omega\abs{g}^{-1/2}$, which leads to UG at the level of the action, we get that the Ricci scalar is still fixed by the coupling constant entering the action when we perform it at the level of the equations of motion
\begin{align}
    R [ \boldsymbol{\tilde{g}} ] = \frac{2\dimM}{\dimM-2}\Lambda\, . 
\end{align}
Hence, we conclude that this gauge fixing is illegal. 

\subsection{Making UG gauge invariant under longitudinal diffeomorphisms}
\label{Subsec:UG_LDiffs}

Let us now consider UG, and let us make it invariant under the full $\mathrm{Diff}$ (i.e., including longitudinal ones). For such a purpose, we will introduce the Stueckelberg fields $Y^a(x)$ as follows:
\begin{widetext}
\begin{align}
    S_{\text{UG-St}}[\boldsymbol{g},Y; \boldsymbol{\omega}]  = \int  {\dd}^\dimM x \omega (x) R \left[ \left( \frac{\omega(x)^2}{\abs{g(Y(x))}}\right)^{\frac{1}{\dimM}} \abs{\det(\frac{\partial Y^a (x)}{\partial x^b})}^{-2/\dimM} g_{cd} (Y(x)) \frac{\partial Y^c (x)}{\partial x^a} \frac{\partial Y^d(x)}{\partial x^b} \right].
\end{align}
\end{widetext}
This can be obtained from the UG action \eqref{eq:actionUG} by making the following replacement everywhere:
\begin{align}
    g_{ab}(x) \rightarrow G_{ab} (Y(x)) =  \frac{\partial Y^c (x)}{\partial x^a} \frac{\partial Y^d(x)}{\partial x^b} g_{cd} (Y(x)),
\end{align}
The fields $Y^a$ are assumed to be invertible functions. The theory enjoys a symmetry that is $\mathrm{WTDiff} \times \mathrm{Diff}'$. Here, $\text{WTDiff}$ is the symmetry that UG already displays, realized in the metric as in Eq.~\eqref{Eq:TDiffs} while leaving the $Y^a$ untouched, and $ \mathrm{Diff}'$ comes from the introduction of the Stueckelberg fields $Y^a$. The latter corresponds to the following realization of the whole set of diffeomorphisms:
\begin{align}
    g_{ab} (x) &\xrightarrow{\mathrm{Diff}'}  \frac{\partial f^c}{\partial x^a} \frac{\partial f^d}{\partial x^b} g_{cd} (f(x)), \\
    \omega(x) &\xrightarrow{\mathrm{Diff}'} \omega(x), \\
    Y^a(x) &\xrightarrow{\mathrm{Diff}'} (f^{-1})^a ( Y(x) ), \label{eq:transfY}
\end{align}
where the transformations of $Y^a$ have been implemented to leave $G_{ab}(Y(x))$  invariant under these transformations:
\begin{align}
    G_{ab} (Y(x)) &\xrightarrow{\mathrm{Diff}'} G_{ab} (Y(x)). 
\end{align}
To recover the original UG action, one just has to take the gauge fixing $Y^a(x)= x^a$. This can always be achieved, due to the invertibility of $Y^a$, by just performing the diffeomorphism for which $f^a = Y^a$, so that in this gauge we find
\begin{equation}
    G_{ab}(Y(x))\Big|_{Y^a(x)=x^a} = g_{ab}(x).
\label{eq:Gtogfix}
\end{equation}
Let us now derive the equations of motion. If we introduce the abbreviation
\begin{align}
    \tilde{G}_{ab}(Y(x)) & :=\left( \frac{\omega(x)}{\abs{G(Y(x))}}\right)^{\frac{1}{\dimM}}G_{ab}(Y(x)), \nonumber \\
    G & :=\det(G_{ab}).
\end{align}
(notice that the $Y$ does not enter the dependency of $\omega$), we realize that the dynamical fields $g^{ab}$ and $Y^a$ only appear in the action through the combination $\tilde{G}^{ab}$, so
\begin{align}
    \delta S_{\text{UG-St}} = \int {\dd}^\dimM x \, \omega(x) \,\left[ R_{ab}[\boldsymbol{\tilde{G}}] - \frac{1}{\dimM} \tilde{G}_{ab} R[\boldsymbol{\tilde{G}}] \right] \delta G^{ab} \Bigg\vert_{Y(x)}, \label{eq:prevarUGsT}
\end{align}
where we took into account that, since $\tilde{G}^{ab}(Y(x))$ has a fixed determinant (equal to $1/\omega(x)$), varying with respect to it is equivalent to varying with respect to the traceless part of $G^{ab}(Y(x))$. In \eqref{eq:prevarUGsT}, the subscript $Y(x)$ indicates that the object in the square bracket is evaluated in $Y(x)$. Now we can compute the variation $G^{ab}$ coming from both $Y^a$ and $g^{ab}$. It is convenient to introduce
\begin{equation}
    M^a{}_b := \partial_b Y^{a},\qquad N^a{}_b := \partial_b (Y^{-1})^{a}
\end{equation}
where $\partial_a$ just means partial with respect to the $a$th slot. These matrices are the inverse of each other in the following sense $M^a{}_b (x)N^b{}_c (Y(x))=N^a{}_b (Y(x)) M^b{}_c (x) =\delta^a{}_c$, as a consequence of the chain rule, which implies
\begin{equation}
    \delta N^b{}_d (Y(x)) = - N^b{}_e (Y(x))\, N^f{}_d(Y(x))\, \delta M^e{}_f (x) \,.\label{eq:MMprop}
\end{equation}
Then: 
\begin{align}
    \delta G^{ab}(Y(x)) &= \delta \left[ g^{cd} N^a{}_c N^b{}_d \right]_{Y(x)}\nonumber\\
    &= - 2 \left[g^{cd} N^a{}_c N^b{}_e N^f{}_d\right]_{Y(x)} \delta M^e{}_f(x) \nonumber\\
    &\quad +\left[\delta g^{cd} N^b{}_c N^b{}_d\right]_{Y(x)}\,, \label{eq:varG}
\end{align}
where we made use of \eqref{eq:MMprop}. The first piece of \eqref{eq:varG}, which corresponds to the variations with respect to $g^{ab}$, leads to:
\begin{align}
    \left[N^a{}_c N^b{}_d\left(R_{ab}[\boldsymbol{\tilde{G}}] - \frac{1}{n} \tilde{G}_{ab} R[\boldsymbol{\tilde{G}}]\right)\right]_{Y(x)}=0, 
\end{align}
which can simply be written as the traceless Einstein equations by multiplying by appropriate factors of $M^a{}_b(x)$:
\begin{align}
    \left[R_{ab}[\boldsymbol{\tilde{G}}] - \frac{1}{n} \tilde{G}_{ab} R[\boldsymbol{\tilde{G}}]\right]_{Y(x)}=0, \label{eq:UGStgeq}
\end{align}
On the other hand, the second term in \eqref{eq:varG} (the equation of motion of $Y^a$), which contains $\delta M^e{}_f(x) = \partial_f \delta Y^e (x)$, can be integrated by parts to obtain
\begin{align}
    &\partial_{d}\bigg\{ \omega(x) \bigg[  2 g^{cf}  N^d{}_f N^a{}_c N^b{}_e \nonumber\\
    &\qquad\qquad \times \bigg(  R_{ab}[\boldsymbol{\tilde{G}}] - \frac{1}{\dimM} \tilde{G}_{ab} R[\boldsymbol{\tilde{G}}]  \bigg) \bigg]_{Y(x)}\bigg\}=0 . \label{eq:UGStYeq}
\end{align}
This equation is redundant because it is an immediate consequence of the first one \eqref{eq:UGStgeq}. For this to be a well-defined Stueckelberg-ization of the UG theory, we have to check that the gauge fixing $Y^a(x)=x^a$, which leads to the UG action, also works at the level of equations of motion. Since the equation of $Y^a$ is redundant, we can just focus on the equation of the metric \eqref{eq:UGStgeq}, which clearly reduces to that of UG if we take $Y^a(x)=x^a$,  thanks to \eqref{eq:Gtogfix}.

We can now try to fix a gauge in which we reach GR. For that purpose, we would make a Weyl transformation \eqref{Eq:Weyl_Resc} that allows us to reach the gauge in which
\begin{align}
    & \left( \frac{\omega(x)}{\abs{g(Y(x))}}\right)^{\frac{1}{\dimM}} \abs{\det(M^a{}_b)}^{-2/\dimM}=1 \nonumber \\
     \Leftrightarrow &\quad
    \omega(x) = \sqrt{\abs{g(Y(x))}}\ \abs{\det(M^a{}_b)} .
\end{align}
In this gauge, the action reduces to
\begin{widetext}
\begin{align}
    \int  {\dd}^\dimM x \ \abs{\det(\frac{\partial Y^a(x)}{\partial x^b})} \sqrt{\abs{g(Y(x))}} \ R \left[  g_{cd} (Y(x)) \frac{\partial Y^c (x)}{\partial x^a} \frac{\partial Y^d(x)}{\partial x^b} \right].
\end{align}
\end{widetext}
and this is the GR action after applying an active diffeomorphism $x^a\to Y^a(x)$. Of course, this gauge transformation is illegal, in the sense that when performed at the level of the action does not lead to traceless equations of motion, unlike what we find in the equations of motion after fixing the gauge in them.


\section{Trinity formulation of GR and UG}
\label{Sec:Trinity}

For the sake of completeness and also to extend and clarify a bit the results presented in~\cite{Nakayama2022}, in this section we will present the trinity formulation of UG. We will also show how the trinity formulation of UG is different from the trinity formulation of GR in the same sense that UG is different from GR: due to the difference in the behavior of the cosmological constant. For that purpose, we will argue that introducing Stueckelberg fields in the trinity formulation of GR still leads to theories that are not inequivalent to the trinity formulation of UG, since we miss the global degree of freedom encoded in the cosmological constant. For this section, we do not perform the analysis of making the trinity formulation of UG invariant under diffeomorphisms because it is more convoluted and it does not add anything new into the discussion.

To present UG and its trinity formulation, let us begin with unimodular gravity in the second order formalism: 
\begin{align}
    S_{\text{UG}_{(2)}} [\boldsymbol{g} ;\boldsymbol{\omega}]= \frac{1}{2 \kappa^2} \int \dd^\dimM x \omega R  [\boldsymbol{\tilde{g}}] + \text{GBH},
    \label{Eq:UG_V2}
\end{align}
where GBH represents the Gibbons-Hawking-York term which is there to ensure that we have a well-defined variational problem~\cite{York1972,GibbonsHawking1977}.

First of all, let us show that this action is equivalent to its version in which we consider an arbitrary connection that we impose to be torsionless and metric compatible, resulting in the Levi-Civita connection. For that purpose, let us consider the following action: 
\begin{widetext}
\begin{align}
    S_{\text{UG}_{(1)}} [\boldsymbol{g} , \boldsymbol{\Gamma} ;\boldsymbol{\omega}] = \frac{1}{2 \kappa^2} \int \dd^\dimM x \omega \Tilde{g}^{a b} R_{a b}[\boldsymbol{\Gamma}]  + \int\dd^\dimM x   \omega \lambda_{a}{}^{b c} T^{a}{}_{b c} \left[  \boldsymbol{\Gamma}  \right] + \int\dd^\dimM x   \omega \hat{\lambda}^{abc} Q_{ab c} \left[  \boldsymbol{g}, \boldsymbol{\Gamma}; \boldsymbol{\omega}  \right] , 
\label{Eq:UG_V1}
\end{align}
\end{widetext}
where we have introduced the torsion $T^{a}{}_{b c} := 2 \Gamma^{a}{}_{[b c]}$ and the nonmetricity $Q_{cab} :=  \nabla^{\Gamma}_{c} \tilde{g}_{ab}$ ($\nabla^{\Gamma}$ being the covariant derivative of $\boldsymbol{\Gamma}$). Notice that we are defining the nonmetricity with respect to the auxiliary metric $\boldsymbol{\Tilde{g}}$, not the dynamical metric $\boldsymbol{g}$. In this way we ensure that the action is Weyl invariant since it is built out from Weyl-invariant objects. The fields $\lambda_{a}{}^{b c}$ and $\hat{\lambda}^{abc}$ are Lagrange multiplier densities that enforce the torsionless and metric-compatibility conditions. These constraints can be solved and plugged back into the action. This automatically leads to the action from Eq.~\eqref{Eq:UG_V2}.

\paragraph*{\textbf{Stueckelberg-ing Palatini equivalent of GR to have Weyl invariance.}}

In the same vein that we did for GR, we can introduce a Stueckelberg field that realizes the Weyl invariance. We want to make the theory invariant under transformations of the form
\begin{align}
    g_{ab} \rightarrow\ee^{2 \phi(x)} g_{ab}, 
\end{align}
which are Weyl transformations realized in the standard way. For such a purpose, we would need to introduce everywhere a Stueckelberg field that compensates the transformation on $g_{ab} (x)$, i.e., a field $\pi(x)$ transforming as 
\begin{align}
    \pi(x) \rightarrow \pi(x) - \phi (x),
\end{align}
that we introduce replacing $g_{ab}$ everywhere in the action by $\ee^{2 \pi} g_{ab}$. In that sense, we consider the theory 
\begin{widetext}
\begin{align}
    S_{\text{GR}_{(1)}} [\boldsymbol{g} , \boldsymbol{\Gamma}] &= \frac{1}{2 \kappa^2} \int \dd^\dimM x \sqrt{\abs{g}} g^{ab} R_{ab}[\boldsymbol{g}, \boldsymbol{\Gamma}]  + \int\dd^\dimM x   \sqrt{\abs{g}} \lambda_{a}{}^{bc} T^{a}{}_{bc} \left[  \boldsymbol{\Gamma}  \right] + \int\dd^\dimM x  \sqrt{\abs{g}} \hat{\lambda}^{abc} Q_{abc} \left[  \boldsymbol{g}, \boldsymbol{\Gamma} \right] , 
\label{Eq:GR_Palatini}
\end{align}
\end{widetext}
This theory is clearly invariant under diffeomorphisms. If we introduce the Stueckelberg field, we promote it to a theory that is Weyl and Diff invariant. The resulting action reads
\begin{align}
    S_{\text{GR-St}_{(1)}} [\boldsymbol{g} , \boldsymbol{\Gamma}, \pi] = S_{\text{GR}_{(1)}} [\ee^{2 \pi} \boldsymbol{g} , \boldsymbol{\Gamma}].
\end{align}
We can now repeat the same exercise that we did before and integrate the constraints. The torsion is trivially integrated, and we simply get that the connection needs to be symmetric. The nonmetricity is a little bit subtle, due to the term $\pi$ appearing in the constraint
\begin{align}
    \nabla_{c}^{\Gamma} \left(\ee^{2 \pi} g_{ab} \right) = 0.
\end{align}
This implies that the connection is compatible with the metric $\ee^{2 \pi}g_{ab}$ (not only the metric, $g_{ab}$). We can plug this back into the action and the result is the Stueckelberg-ized version of GR that is invariant under diffeomorphisms and Weyl transformations. The inequivalence between this version of GR and UG has been shown in Sec. \ref{Subsec:GR_Weyl} and it arises here also. Hence, the Palatini version of UG is again different from the Palatini version of GR, in the sense that there is an additional global degree of freedom.

\subsection{Metric teleparallel equivalent to UG}
\label{Subsec:Metric_Teleparallel_UG}

We consider the most general even-parity scalar that can be built with the torsion that is given by
\begin{align}
    \mathbb{T} [ \boldsymbol{g},\boldsymbol{\Gamma};\boldsymbol{\omega}] := -\frac{c_1}{4} T_{abc} T^{abc} - \frac{c_2}{2} T_{bac} T^{abc} + c_3 T^b{}_{ab} T^{ca}{}_c, 
\end{align}
where indices have been raised/lowered with $\boldsymbol{\tilde{g}}$. We consider the most general action involving this term and impose that the Riemann curvature is zero and compatibility of the connection with the auxiliary metric $\boldsymbol{\tilde{g}}$. This leads to the following Lagrangian
\begin{widetext}
\begin{align}
    S_{\mathbb{T}} [ \boldsymbol{g}, \boldsymbol{\Gamma}; \boldsymbol{\omega}] = \frac{1}{2 \kappa^2} \int \dd^\dimM x \omega \mathbb{T} [\boldsymbol{g},\boldsymbol{\Gamma};\boldsymbol{\omega}]  +  
 \int\dd^\dimM x   \omega \lambda^{b c d}{}_a R_{bcd}{}^{a} [  \boldsymbol{\Gamma}  ]  + \int \dd^\dimM x  \omega \hat{\lambda}^{ab c} Q_{ab c} [  \boldsymbol{g}, \boldsymbol{\Gamma} ; \boldsymbol{\omega} ] .
\end{align}
\end{widetext}
Let us begin again by solving the constraints. First of all, we have the constraint that the connection is locally flat:
\begin{align}
    R_{a b c }{}^{d} \left[  \boldsymbol{\Gamma}  \right] = 0.
\end{align}
Following~\cite{BeltranJimenez2018}, we see that the most general connection which has zero Riemann curvature is the one associated with an arbitrary matrix function belonging to the general linear group $GL(4,\mathbb{R})$ since the curvature is the field strength associated with the connection. Hence, since the trivial connection (all components vanishing) satisfies the constraint, any $GL(4,\mathbb{R})$ transformation of the trivial connection will lead to another solution:
\begin{align}
    \Gamma^a{}_{ b c }{} = \left( \Lambda^{-1} \right)^{a}{}_{d } \partial_{b} \Lambda^{d}{}_{c}. 
    \label{Eq:TrivialConnection}
\end{align}
If we now impose the second constraint, we have:
\begin{align}
    \nabla^{\Gamma}_{a} \tilde{g}_{b c} = \partial_{a} \tilde{g}_{bc} -  \Gamma^{d}{}_{ab} \tilde{g}_{ d c} -   \Gamma^{d}{}_{ac} \tilde{g}_{ b d} =0.
\end{align}
Plugging the general connection from Eq.~\eqref{Eq:TrivialConnection}, we find the constraint
\begin{align}
    \partial_{a} \tilde{g}_{bc} = \tilde{g}_{d c} \left( \Lambda^{-1} \right)^{d}{}_{e} \partial_{a} \Lambda^{e}{}_{b} + \tilde{g}_{b d } \left( \Lambda^{-1} \right)^{d}{}_{e} \partial_{a} \Lambda^{e}{}_{c}. 
\end{align}
The torsion is the antisymmetric part of the connection, which for the connection in Eq.~\eqref{Eq:TrivialConnection} is given by:
\begin{align}
    T^a{}_{b c} = 2 \left( \Lambda^{-1} \right)^{a}{}_{d} \partial_{[b} \Lambda^{d}{}_{c]}. 
\end{align}
In the absence of nonmetricity, we can relate the Ricci scalar $R[\boldsymbol{\Gamma}]$ associated with the general connection $\boldsymbol{\Gamma}$ and the Ricci scalar $R[\boldsymbol{\tilde{g}}]$ associated with the Levi-Civita connection compatible with $\boldsymbol{\tilde{g}}$ as:
\begin{align}           
   R[\boldsymbol{\tilde{g}},\boldsymbol{\Gamma}]  = R[\boldsymbol{\tilde{g}}] - \mathbb{T}[ \boldsymbol{g},\boldsymbol{\Gamma};\boldsymbol{\omega}]_{c_i=1} + 2 \tilde{\nabla}_{d} T^{d}.
\end{align}
Note that the scalar $\mathbb{T}$ with $c_1 = c_2 = c_3 = 1$ is equivalent to the Ricci scalar of the Levi-Civita connection, which constitutes the UG Lagrangian. Therefore, this formulation of UG is equivalent to the original formulation of UG, that in Eq.~\eqref{Eq:UG_V2}, and hence is different from GR.

\subsection{Symmetric teleparallel equivalent to UG}
\label{Subsec:Symmetric_Teleparallel_UG}
We want to describe now everything in terms of the nonmetricity. For that purpose, we introduce the most general parity-even scalar that is quadratic in the nonmetricity
\begin{widetext}
\begin{align}
    \mathbb{Q} [  \boldsymbol{g}, \boldsymbol{\Gamma} ; \boldsymbol{\omega}] := - \frac{c_1}{4} Q_{abc} Q^{abc} + \frac{c_2}{2} Q_{abc} Q^{bac} + \frac{c_3}{4} Q_{ab}{}^b Q^{ac}{}_c + (c_4 - 1) Q^b{}_{ba} Q_c{}^{ca} - \frac{c_5}{2} Q_{ab}{}^b Q_c{}^{ca},
\end{align}
\end{widetext}
To this action we have to incorporate the Lagrange multipliers that enforce the zero Riemann curvature condition and restrict the connection to be torsionless: 
\begin{widetext}
\begin{align}
    S_{\mathbb{Q}} [\boldsymbol{g}, \boldsymbol{\Gamma}; \boldsymbol{\omega}] = \frac{1}{2 \kappa^2} \int \dd^\dimM x \omega  \mathbb{Q} [\boldsymbol{g}, \boldsymbol{\Gamma} ; \boldsymbol{\omega}] +  
 \int \dd^\dimM x   \omega \lambda^{b c d}{}_a R_{b c d}{}^{a}  \left[  \boldsymbol{\Gamma}  \right]  + \int \dd^\dimM x  \omega \lambda_{a}{}^{b c} T^a{}_{b c} [\boldsymbol{\Gamma}] .
\end{align}
\end{widetext}
For a torsionless connection, we have the following relation among the Ricci scalar for the general connection and the scalar $\mathbb{Q}$ with suitable choices of the parameters $c_i$: 
\begin{align}
    R[\boldsymbol{\tilde{g}},\boldsymbol{\Gamma}] = R[\boldsymbol{\tilde{g}}] - \mathbb{Q}  [ \boldsymbol{g}, \boldsymbol{\Gamma} ; \boldsymbol{\omega}]_{c_i=1} +  \tilde{\nabla}_{a} \left( Q^{ab}{}_b - Q_b{}^{ba} \right).
\end{align}
Thus, the dynamics encoded by the tensor $\mathbb{Q}$ for this choice of parameters is the same as the Ricci scalar for the metric $\boldsymbol{\tilde{g}}$ for a flat connection: $R[\boldsymbol{\tilde{g}},\boldsymbol{\Gamma}]  = 0$. Again, we  find the same result that we have found for the metric teleparallel equivalent. Introducing the Stueckelberg fields would not alter this conclusion in any way, and hence we would obtain a theory invariant under Diffs and Weyl transformations. However, the global degree of freedom that we miss in the GR version of the trinity formulation would still be lacking in the Stueckelberg-ized version. 

Up to this point, we have carefully presented the trinity formulation of UG and we have compared it to the trinity formulation of GR. As we have advanced, there is a mismatch between the global degrees of freedom: the UG trinity displays the cosmological constant whereas the GR trinity contains no degrees of freedom. Hence, we also conclude that the trinity formulation of UG is inequivalent to GR (and its trinity formulation) because of the global degree of freedom present in the theory, a fact that does not seem to be emphasized enough in~\cite{Nakayama2022}.

\section{Conclusions}
\label{Sec:Conclusions}

In this paper we have explored the relation between GR and UG from the point of view of their gauge symmetries. We have posed the question of whether there exists a parent theory that is invariant under Weyl transformations and diffeomorphisms such that GR and UG are suitable gauge fixings of it. We approached the problem by introducing Stueckelberg fields, in GR to make it invariant under Weyl transformations and in UG to make it invariant under the whole set of diffeomorphisms. We have found that the resulting theories are not equivalent, and we have isolated the obstruction that forbids one to find such a theory: the cosmological constant that appears in UG as a global degree of freedom, something that is tightly related to the existence of a background volume form.

The difference between UG and GR points toward future extensions of UG that may lead to an interesting phenomenology. The cosmological constant, as it appears in UG as a global degree of freedom, contains no dynamics. However, it is clear from our analysis that future works attempting to make extensions of UG should go in the direction of giving a dynamics to the cosmological constant~\cite{Perez2017,Alexander2018}. It is this approach that can give rise to a phenomenology that is different from the phenomenology of GR as it is found in~\cite{Perez2017}. Direct extensions of UG in the form of higher derivative generalizations or additional fields, lead to theories that are equivalent to their associated higher derivative generalizations from GR~\cite{Carballo-Rubio2022}.

We have also taken the opportunity to present a careful discussion of the trinity formulation of UG and clarify its relation with the trinity formulation of GR. We have developed some points regarding the inequivalence between these formulations that were not completely clear in~\cite{Nakayama2022}.

\begin{acknowledgments}
The authors thank Jose Beltr\'an Jim\'enez for useful discussions, contributions to this research and for his careful revision of the manuscript and feedback. The authors also thank Carlos Barcel\'o, Luis J. Garay and Ra\'ul Carballo-Rubio for enlightening discussions and valuable feedback, Miguel S\'anchez for helpful discussions on the relation between the group of diffeomorphisms and its Lie algebra and Roberto Percacci for useful correspondence. The authors are also grateful to the anonymous referee for their useful suggestions that have allowed us to clarify the presentation of the manuscript. Financial support was provided by the Spanish Government through the projects PID2020-118159GB-C43, PID2020-118159GB-C44, and by the Junta de Andaluc\'{\i}a through the project FQM219. GGM is funded by the Spanish Government fellowship FPU20/01684 and acknowledges financial support from the Severo Ochoa grant CEX2021-001131-S funded by MCIN/AEI/10.13039/501100011033. .J.C. was supported by the European Regional Development Fund through the Mobilitas Pluss postdoctoral Grant No.MOBJD1035, as well as by the Center of Excellence TK133 ``The Dark Side of the Universe'' and the project PRG356 funded by the Estonian Research Council.
\end{acknowledgments}

\appendix 

\section{Examples of illegal gauge fixings}
\label{Sec:Illegal_gauge}

Let us study two examples of theories in which performing a gauge fixing at the level of the action and at the level of the equations of motion do not lead to the same results. This is what we will call an illegal gauge fixing. Let us emphasize again that an illegal gauge fixing always refers to imposing the conditions on the fields before performing the variation on the action. By no means we are suggesting that it is not possible to use these gauges in the equations of motion. Notice, however, that it is also possible to impose them at the action level as long as one introduces suitable Lagrange multipliers to enforce the constraints (the gauge fixing conditions). Their variation leads to a set of algebraic equations equivalent to the equations found by performing the gauge fixing directly in the equations of motion.

\subsection{Electrodynamics in the Coulomb gauge with matter content}

Take Maxwell action coupled to a conserved current $j^{a}$: 

\begin{align}
    S =   \int \dd^{\dimM} x \left[ - \frac{1}{4} F^{ab} F_{ab} + A_{a} j^{a} \right].
\end{align}
Its equations of motion can be directly computed and one finds
\begin{align}
    \Box A^{a} - \partial^{a} ( \partial_{b} A^{b} ) = j^{a}. 
\end{align}
Now we can always perform a gauge transformation such that $A_0 = 0$. To see this, under a gauge transformation we have: 
\begin{align}
    A_{a} \rightarrow A'_{a} = A_{a} + \partial_{a} \alpha,
\end{align}
and we can always choose an $\alpha$ such that $A'_0 = 0$; for example,
\begin{align}
    \alpha(t,\boldsymbol{x}) = - \int_{t_0}^t \dd t' A_0(t', \boldsymbol{x})
\end{align}
does the job. In this gauge $A^{a} \partial_{a}$ is a purely spatial vector that we denote as $\boldsymbol{A}$, and the equations of motion reduce to
\begin{align}
    & \partial_t \left( \nabla \cdot \boldsymbol{A} \right) = j^0, \label{eq:Gausslaw}\\
    & (- \partial_t^2 + \nabla^2 ) \boldsymbol{A} - \nabla ( \nabla \cdot \boldsymbol{A} ) = \boldsymbol{j}.
\end{align}
The same is not obtained if we fix the gauge at the level of the action though. In this sense (i.e., at the action level), this gauge fixing is ``illegal''. To see this, upon substituting $A_0 = 0$ in the action we find
\begin{align}
    S\big|_{A^0=0} = \int \dd^n x \left[\frac{1}{2} ( \partial_t \boldsymbol{A} )^2 - \frac{1}{2} ( \nabla \times \boldsymbol{A} )^2 + \boldsymbol{j} \cdot \boldsymbol{A} \right].
\end{align}
If we vary this action, we notice that all the information related to charge conservation (the Gauss law \eqref{eq:Gausslaw}) is lost, and we only get the second equation for $\boldsymbol{A}$: 
\begin{align}
    (- \partial_t^2 + \nabla^2 ) \boldsymbol{A} - \nabla ( \nabla \cdot \boldsymbol{A} ) = \boldsymbol{j}. 
\end{align}
Thus, this gauge fixing cannot be performed at the level of the action in order to reduce the number of variables since we lose information about the existence of a conserved charge. Although it should be possible to still fix this gauge at the level of the action and implement a variational principle (by enforcing the Gauss law somehow), we will still refer to these kinds of gauge fixings as illegal gauge fixings.
\subsection{Stueckelberg-ing Proca and the Lorenz gauge}
Let us consider now a massive spin-1 theory:
\begin{align}
    S_{\text{Proca}} = \int \dd^n x \left[ - \frac{1}{4} F_{ab} F^{ab} - \frac{1}{2} m^2A_{a}A^{a} \right]. 
\end{align}
We introduce a Stueckelberg field by making the replacement $A_{a} \rightarrow A_{a} + \partial_{a} \varphi$, giving the following action: 
\begin{align}
    S_{\text{Proca-St}} & = \int \dd^n x \Big[ - \frac{1}{4} F_{ab} F^{ab} - \frac{1}{2} m^2 A_{a} A^{a}  \nonumber  \\
    &\quad - m^2 A^{a} \partial_{a} \varphi - \frac{1}{2} m^2 \partial_{a} \varphi \partial^{a} \varphi \Big]. 
\end{align}
This action is invariant under the following gauge transformation that hits the two fields
\begin{align}
    A_a &\to A^\prime_a=A_a +\partial_{a}\alpha\, \\
    \varphi &\to \varphi^\prime=\varphi -\alpha\,.
\end{align}
The equations of motion of the theory are
\begin{align}
    & \Box \varphi + \partial_{a} A^{a} = 0, \\
    & \partial_{a} F^{ab} - m^2 A^{b} - m^2 \partial^{b} \varphi = 0. 
\end{align} 
We now can fix the gauge in which $\partial_{a} A^{a} = 0$, which is the Lorenz gauge since we can always find a gauge transformation that ends up in that gauge. To see this, given a field configuration $A_{a}$, upon a gauge transformation we have that the divergence of the transformed field is:
\begin{align}
    \partial_{a} A^{ \prime a} =  \partial_{a} A^{a} + \Box \alpha,  
\end{align}
Hence, we can always find a $\alpha$ obeying the following equation:
\begin{align}
    \Box \alpha = - \partial_{a} A^{a}. 
\end{align}

The situation is completely different if we try to implement this gauge fixing at the level of the action. The whole coupling between the $A^{a}$ and the Stueckelberg field occurs through a term that vanishes in the Lorenz gauge. To see it explicitly, we can take the action and perform an integration by parts to reach
\begin{align}
     S_{\text{Proca-St}} = &  \int \dd^n x \Big[ - \frac{1}{4} F_{ab} F^{ab} - \frac{1}{2} m^2 A_{a} A^{a} \nonumber \\ 
     & + m^2  \partial_{a} A^{a} \varphi - \frac{1}{2} m^2 \partial_{a} \varphi \partial^{a} \varphi \Big]. 
\end{align}
If we fix the Lorenz gauge, we lose the coupling between $A^{a}$ and $\varphi$, reaching the following action
\begin{align}
    S_{\text{Proca-St}}\big|_{\partial_{a} A^a =0} & = \int \dd^n x \Big[ - \frac{1}{4} F_{ab} F^{ab} \nonumber \\
    & \quad- \frac{1}{2} m^2 A_{a} A^{a}  - \frac{1}{2} m^2 \partial_{a} \varphi \partial^{a} \varphi \Big],
\end{align}
and its equations of motion are clearly the decoupled equations of the vector $A^{a}$ and the scalar $\varphi$: 
\begin{align}
    & \Box \varphi = 0, \\
    & \partial_{a} F^{ab} - m^2 A^{b}  = 0. 
\end{align} 
We conclude that again, this (partial) gauge fixing is not acceptable since it does not lead to the same equations of motion.

\section{Passive and active diffeomorphisms}
\label{sec:diffs}

In this appendix we discuss the difference between passive and active diffeomorphisms in the presence of background structures, in order to fix terminology and be more clearer in the subsequent sections. In a sense, as groups, both active and passive diffeomorphisms are the same group but simply realized in a different way on the fields.

On the one hand, \emph{passive diffeomorphisms} correspond to general coordinate transformations (these we are abbreviating in this paper as $\mathrm{GCT}$). These transformations only relabel the points of the manifold with new coordinates but do not move them nor the fields. For a given tensor field, a coordinate transformation affects both its components and its functional dependency on the coordinates. In practice, a coordinate transformation $x^a\to 
 x'{}^a=f^a(x)$ modifies a tensor field $T^{a...}{}_{b...}$, as it is well-known, as
 \begin{align}
     T^{a...}{}_{b...} (x) = & (J^{-1})^a{}_c\big|_{f(x)}...\, J^d{}_b\big|_x...\, T'{}^{c...}{}_{d...}(f(x)), \nonumber \\
     J^c{}_a\big|_x := & \frac{\partial f^c}{\partial x^a}(x)\,.
 \end{align}
At the level of the action it is also important to keep in mind how the product of  $\dd^\dimM x$ and a volume scalar density transforms under these transformations. They do it in an opposite way so  their product is invariant:
\begin{equation}
    \mathfrak{s}(x) \dd^\dimM x = \mathfrak{s}'(f(x)) \dd^\dimM x' .\label{eq:GCTsddx}
\end{equation}

On the other hand, \emph{active diffeomorphisms} truly act on the points of the manifold (abbreviated as $\mathrm{Diff}$). With respect to a fixed coordinate system, the fields must transform so that they take the same value at the same points. In practice one does the substitution:\footnote{
    Infinitesimally, the effect of an active diffeomorphism is described by the Lie derivative $\delta_{\xi} T^{a...}{}_{b...} = \left( \mathcal{L}_{\xi} T \right)^{a...}{}_{b...}$. In particular, for the metric field, we have the following infinitesimal transformation $
    \delta_{\xi} g_{ab} = \left( \mathcal{L}_{\xi} g  \right)_{ab} = \nabla_{a} \xi_{b} + \nabla_{b} \xi_{a}.$
}
\begin{align}
    T^{a...}{}_{b...} (x) \xrightarrow{\mathrm{Diff}}  (J^{-1})^a{}_c\big|_{f(x)}... \,J^d{}_b\big|_x...\,T^{c...}{}_{d...}(f(x))\,, \label{eq:activediff}
\end{align}
where now in $x^a \to x'{}^a=f^a(x)$, $f^a$ should not be seen as functions describing ``how the coordinates of an arbitrary given point change'' but as the functions that, at a fixed set of coordinates, give ``the new coordinates of the considered point after the transformation.'' Of course, an additional factor with a power of the determinant $\abs{J}$ must be added to \eqref{eq:activediff} in the case of a tensor density. By default, the functional expressions of all the fields are assumed to be affected by these $\mathrm{Diff}$. 

In general, any theory written in a covariant way (i.e., in abstract tensor notation) is invariant under both $\mathrm{GCT}$ and also under $\mathrm{Diff}$ that hit all the of the tensorial quantities \emph{simultaneously}. Because of~\eqref{eq:GCTsddx} the invariance of a covariant action under $\mathrm{GCT}$ is guaranteed. To see that such an action is also invariant under  $\mathrm{Diff}$ we first perform the substitutions \eqref{eq:subst} for \emph{all} tensors and densities but not for the ${\dd}^\dimM x$, which remains the same (because coordinates are not changing in an active diffeomorphism). As a result we get (where $\mathfrak{s}$ represent some scalar density and $D$ a certain domain in $\mathbb{R}^\dimM$)
\begin{align}
    S_0:= & \int_{D}  \mathcal{L}(x) \,  \mathfrak{s}(x) \, {\dd}^\dimM x \nonumber \\
    & \quad\xrightarrow{\mathrm{Diff}} \quad\int_{f(D)}  \mathcal{L}(f(x)) \, \Big(\mathfrak{s}(f(x)) \abs{J}\Big)
    {\dd}^\dimM x \label{eq:subst}
\end{align}
(if done with care for all the fields, the Lagrangian $\mathcal{L}$ should transform exactly as in~\eqref{eq:subst}). Finally, we perform a change in the variable of integration and work with $y^a=f^a(x)$ instead of $x^a$. Then, from the $ {\dd}^\dimM x$ we will get an inverse Jacobian $\abs{J}^{-1}$ so at the end, the action looks as
\begin{equation}
     \int_{D} \mathcal{L}(y) \,  \mathfrak{s}(y) \, {\dd}^\dimM y
\end{equation}
which is nothing but the original action $S_0$ because $y$ is just a dummy variable.

Although mathematically different, the transformations $\mathrm{GCT}$ and  $\mathrm{Diff}$ are actually in one-to-one correspondence. From now on we will adopt the passive point of view and simply refer to this symmetry as $\mathrm{GCT}$.

Besides all of this, the situation changes dramatically when we consider \emph{background structures}. We say that a tensor density $\Omega^{a...}{}_{b...}$ is a background structure if it is not affected by active diffeomorphisms. More precisely, such a field transforms as an scalar (ignoring indices and density weights),
\begin{equation}
    \Omega^{a...}{}_{b...} (x) \quad\xrightarrow{\mathrm{Diff}} \quad  \,\Omega^{a...}{}_{b...}(f(x))\,.\label{eq:bgstructransf}
\end{equation}
When there are background structures, the theory distinguishes between $\mathrm{GCT}$ and active diffeomorphisms (since they preserve $\Omega^{a...}{}_{b...}$ in the sense of \eqref{eq:bgstructransf}).

\makeatletter
\def\bibsection{%
  \par
  \baselineskip26\p@
  \bib@device{\linewidth}{82\p@}%
  \nobreak\@nobreaktrue
  \addvspace{19\p@}%
  \par
}
\makeatother

\bibliography{ug_stueckelberg_biblio}

\begin{thebibliography}{26}%
\makeatletter
\providecommand \@ifxundefined [1]{%
 \@ifx{#1\undefined}
}%
\providecommand \@ifnum [1]{%
 \ifnum #1\expandafter \@firstoftwo
 \else \expandafter \@secondoftwo
 \fi
}%
\providecommand \@ifx [1]{%
 \ifx #1\expandafter \@firstoftwo
 \else \expandafter \@secondoftwo
 \fi
}%
\providecommand \natexlab [1]{#1}%
\providecommand \enquote  [1]{``#1''}%
\providecommand \bibnamefont  [1]{#1}%
\providecommand \bibfnamefont [1]{#1}%
\providecommand \citenamefont [1]{#1}%
\providecommand \href@noop [0]{\@secondoftwo}%
\providecommand \href [0]{\begingroup \@sanitize@url \@href}%
\providecommand \@href[1]{\@@startlink{#1}\@@href}%
\providecommand \@@href[1]{\endgroup#1\@@endlink}%
\providecommand \@sanitize@url [0]{\catcode `\\12\catcode `\$12\catcode
  `\&12\catcode `\#12\catcode `\^12\catcode `\_12\catcode `\%12\relax}%
\providecommand \@@startlink[1]{}%
\providecommand \@@endlink[0]{}%
\providecommand \url  [0]{\begingroup\@sanitize@url \@url }%
\providecommand \@url [1]{\endgroup\@href {#1}{\urlprefix }}%
\providecommand \urlprefix  [0]{URL }%
\providecommand \Eprint [0]{\href }%
\providecommand \doibase [0]{http://dx.doi.org/}%
\providecommand \selectlanguage [0]{\@gobble}%
\providecommand \bibinfo  [0]{\@secondoftwo}%
\providecommand \bibfield  [0]{\@secondoftwo}%
\providecommand \translation [1]{[#1]}%
\providecommand \BibitemOpen [0]{}%
\providecommand \bibitemStop [0]{}%
\providecommand \bibitemNoStop [0]{.\EOS\space}%
\providecommand \EOS [0]{\spacefactor3000\relax}%
\providecommand \BibitemShut  [1]{\csname bibitem#1\endcsname}%
\let\auto@bib@innerbib\@empty
\bibitem [{\citenamefont {Carballo-Rubio}\ \emph {et~al.}(2022)\citenamefont
  {Carballo-Rubio}, \citenamefont {Garay},\ and\ \citenamefont
  {Garc\'\i{}a-Moreno}}]{Carballo-Rubio2022}%
  \BibitemOpen
  \bibfield  {author} {\bibinfo {author} {\bibfnamefont {R.}~\bibnamefont
  {Carballo-Rubio}}, \bibinfo {author} {\bibfnamefont {L.~J.}\ \bibnamefont
  {Garay}}, \ and\ \bibinfo {author} {\bibfnamefont {G.}~\bibnamefont
  {Garc\'\i{}a-Moreno}},\ }\href {\doibase 10.1088/1361-6382/aca386} {\bibfield
   {journal} {\bibinfo  {journal} {Class. Quant. Grav.}\ }\textbf {\bibinfo
  {volume} {39}},\ \bibinfo {pages} {243001} (\bibinfo {year} {2022})},\
  \Eprint {http://arxiv.org/abs/2207.08499} {arXiv:2207.08499 [gr-qc]}
  \BibitemShut {NoStop}%
\bibitem [{\citenamefont {Alvarez}\ and\ \citenamefont
  {Velasco-Aja}(2023)}]{Alvarez2023}%
  \BibitemOpen
  \bibfield  {author} {\bibinfo {author} {\bibfnamefont {E.}~\bibnamefont
  {Alvarez}}\ and\ \bibinfo {author} {\bibfnamefont {E.}~\bibnamefont
  {Velasco-Aja}}\ }(\bibinfo {year} {2023})\ \Eprint
  {http://arxiv.org/abs/2301.07641} {arXiv:2301.07641 [gr-qc]} \BibitemShut
  {NoStop}%
\bibitem [{\citenamefont {Henneaux}\ and\ \citenamefont
  {Teitelboim}(1989)}]{Henneaux1989}%
  \BibitemOpen
  \bibfield  {author} {\bibinfo {author} {\bibfnamefont {M.}~\bibnamefont
  {Henneaux}}\ and\ \bibinfo {author} {\bibfnamefont {C.}~\bibnamefont
  {Teitelboim}},\ }\href {\doibase
  https://doi.org/10.1016/0370-2693(89)91251-3} {\bibfield  {journal} {\bibinfo
   {journal} {Physics Letters B}\ }\textbf {\bibinfo {volume} {222}},\ \bibinfo
  {pages} {195 } (\bibinfo {year} {1989})}\BibitemShut {NoStop}%
\bibitem [{\citenamefont {Garay}\ and\ \citenamefont
  {Garc\'\i{}a-Moreno}(2023)}]{Garay2023}%
  \BibitemOpen
  \bibfield  {author} {\bibinfo {author} {\bibfnamefont {L.~J.}\ \bibnamefont
  {Garay}}\ and\ \bibinfo {author} {\bibfnamefont {G.}~\bibnamefont
  {Garc\'\i{}a-Moreno}},\ }\href {\doibase 10.1007/JHEP03(2023)027} {\bibfield
  {journal} {\bibinfo  {journal} {JHEP}\ }\textbf {\bibinfo {volume} {03}},\
  \bibinfo {pages} {027} (\bibinfo {year} {2023})},\ \Eprint
  {http://arxiv.org/abs/2301.03503} {arXiv:2301.03503 [hep-th]} \BibitemShut
  {NoStop}%
\bibitem [{\citenamefont {\'Alvarez}\ \emph {et~al.}(2016)\citenamefont
  {\'Alvarez}, \citenamefont {Gonzalez-Martin},\ and\ \citenamefont
  {Mart\'in}}]{Alvarez2016}%
  \BibitemOpen
  \bibfield  {author} {\bibinfo {author} {\bibfnamefont {E.}~\bibnamefont
  {\'Alvarez}}, \bibinfo {author} {\bibfnamefont {S.}~\bibnamefont
  {Gonzalez-Martin}}, \ and\ \bibinfo {author} {\bibfnamefont {C.~P.}\
  \bibnamefont {Mart\'in}},\ }\href {\doibase 10.1140/epjc/s10052-016-4384-2}
  {\bibfield  {journal} {\bibinfo  {journal} {Eur. Phys. J. C}\ }\textbf
  {\bibinfo {volume} {76}},\ \bibinfo {pages} {554} (\bibinfo {year} {2016})},\
  \Eprint {http://arxiv.org/abs/1605.02667} {arXiv:1605.02667 [hep-th]}
  \BibitemShut {NoStop}%
\bibitem [{\citenamefont {Carballo-Rubio}\ \emph {et~al.}(2019)\citenamefont
  {Carballo-Rubio}, \citenamefont {Di~Filippo},\ and\ \citenamefont
  {Moynihan}}]{CarballoRubio2019}%
  \BibitemOpen
  \bibfield  {author} {\bibinfo {author} {\bibfnamefont {R.}~\bibnamefont
  {Carballo-Rubio}}, \bibinfo {author} {\bibfnamefont {F.}~\bibnamefont
  {Di~Filippo}}, \ and\ \bibinfo {author} {\bibfnamefont {N.}~\bibnamefont
  {Moynihan}},\ }\href {\doibase 10.1088/1475-7516/2019/10/030} {\bibfield
  {journal} {\bibinfo  {journal} {JCAP}\ }\textbf {\bibinfo {volume} {10}},\
  \bibinfo {pages} {030} (\bibinfo {year} {2019})},\ \Eprint
  {http://arxiv.org/abs/1811.08192} {arXiv:1811.08192 [hep-th]} \BibitemShut
  {NoStop}%
\bibitem [{\citenamefont {Eichhorn}(2013)}]{Eichhorn2013}%
  \BibitemOpen
  \bibfield  {author} {\bibinfo {author} {\bibfnamefont {A.}~\bibnamefont
  {Eichhorn}},\ }\href {\doibase 10.1088/0264-9381/30/11/115016} {\bibfield
  {journal} {\bibinfo  {journal} {Classical and Quantum Gravity}\ }\textbf
  {\bibinfo {volume} {30}},\ \bibinfo {pages} {115016} (\bibinfo {year}
  {2013})}\BibitemShut {NoStop}%
\bibitem [{\citenamefont {Eichhorn}(2015)}]{Eichhorn2015}%
  \BibitemOpen
  \bibfield  {author} {\bibinfo {author} {\bibfnamefont {A.}~\bibnamefont
  {Eichhorn}},\ }\href {\doibase 10.1007/JHEP04(2015)096} {\bibfield  {journal}
  {\bibinfo  {journal} {JHEP}\ }\textbf {\bibinfo {volume} {04}},\ \bibinfo
  {pages} {096} (\bibinfo {year} {2015})},\ \Eprint
  {http://arxiv.org/abs/1501.05848} {arXiv:1501.05848 [gr-qc]} \BibitemShut
  {NoStop}%
\bibitem [{\citenamefont {De~Brito}\ \emph {et~al.}(2019)\citenamefont
  {De~Brito}, \citenamefont {Eichhorn},\ and\ \citenamefont
  {Pereira}}]{deBrito2019}%
  \BibitemOpen
  \bibfield  {author} {\bibinfo {author} {\bibfnamefont {G.~P.}\ \bibnamefont
  {De~Brito}}, \bibinfo {author} {\bibfnamefont {A.}~\bibnamefont {Eichhorn}},
  \ and\ \bibinfo {author} {\bibfnamefont {A.~D.}\ \bibnamefont {Pereira}},\
  }\href {\doibase 10.1007/JHEP09(2019)100} {\bibfield  {journal} {\bibinfo
  {journal} {JHEP}\ }\textbf {\bibinfo {volume} {09}},\ \bibinfo {pages} {100}
  (\bibinfo {year} {2019})},\ \Eprint {http://arxiv.org/abs/1907.11173}
  {arXiv:1907.11173 [hep-th]} \BibitemShut {NoStop}%
\bibitem [{\citenamefont {de~Brito}\ \emph
  {et~al.}(2021{\natexlab{a}})\citenamefont {de~Brito}, \citenamefont
  {Pereira},\ and\ \citenamefont {Vieira}}]{deBrito2020}%
  \BibitemOpen
  \bibfield  {author} {\bibinfo {author} {\bibfnamefont {G.~P.}\ \bibnamefont
  {de~Brito}}, \bibinfo {author} {\bibfnamefont {A.~D.}\ \bibnamefont
  {Pereira}}, \ and\ \bibinfo {author} {\bibfnamefont {A.~F.}\ \bibnamefont
  {Vieira}},\ }\href {\doibase 10.1103/PhysRevD.103.104023} {\bibfield
  {journal} {\bibinfo  {journal} {Phys. Rev. D}\ }\textbf {\bibinfo {volume}
  {103}},\ \bibinfo {pages} {104023} (\bibinfo {year} {2021}{\natexlab{a}})},\
  \Eprint {http://arxiv.org/abs/2012.08904} {arXiv:2012.08904 [hep-th]}
  \BibitemShut {NoStop}%
\bibitem [{\citenamefont {Herrero-Valea}\ and\ \citenamefont
  {Santos-Garcia}(2020)}]{Herrero-Valea2020}%
  \BibitemOpen
  \bibfield  {author} {\bibinfo {author} {\bibfnamefont {M.}~\bibnamefont
  {Herrero-Valea}}\ and\ \bibinfo {author} {\bibfnamefont {R.}~\bibnamefont
  {Santos-Garcia}},\ }\href {\doibase 10.1007/JHEP09(2020)041} {\bibfield
  {journal} {\bibinfo  {journal} {JHEP}\ }\textbf {\bibinfo {volume} {09}},\
  \bibinfo {pages} {041} (\bibinfo {year} {2020})},\ \Eprint
  {http://arxiv.org/abs/2006.06698} {arXiv:2006.06698 [hep-th]} \BibitemShut
  {NoStop}%
\bibitem [{\citenamefont {Percacci}(2018)}]{Percacci2017}%
  \BibitemOpen
  \bibfield  {author} {\bibinfo {author} {\bibfnamefont {R.}~\bibnamefont
  {Percacci}},\ }\href {\doibase 10.1007/s10701-018-0189-5} {\bibfield
  {journal} {\bibinfo  {journal} {Found. Phys.}\ }\textbf {\bibinfo {volume}
  {48}},\ \bibinfo {pages} {1364} (\bibinfo {year} {2018})},\ \Eprint
  {http://arxiv.org/abs/1712.09903} {arXiv:1712.09903 [gr-qc]} \BibitemShut
  {NoStop}%
\bibitem [{\citenamefont {de~Brito}\ \emph
  {et~al.}(2021{\natexlab{b}})\citenamefont {de~Brito}, \citenamefont
  {Melichev}, \citenamefont {Percacci},\ and\ \citenamefont
  {Pereira}}]{deBrito2021}%
  \BibitemOpen
  \bibfield  {author} {\bibinfo {author} {\bibfnamefont {G.~P.}\ \bibnamefont
  {de~Brito}}, \bibinfo {author} {\bibfnamefont {O.}~\bibnamefont {Melichev}},
  \bibinfo {author} {\bibfnamefont {R.}~\bibnamefont {Percacci}}, \ and\
  \bibinfo {author} {\bibfnamefont {A.~D.}\ \bibnamefont {Pereira}},\ }\href
  {\doibase 10.1007/JHEP12(2021)090} {\bibfield  {journal} {\bibinfo  {journal}
  {JHEP}\ }\textbf {\bibinfo {volume} {12}},\ \bibinfo {pages} {090} (\bibinfo
  {year} {2021}{\natexlab{b}})},\ \Eprint {http://arxiv.org/abs/2105.13886}
  {arXiv:2105.13886 [gr-qc]} \BibitemShut {NoStop}%
\bibitem [{\citenamefont {Gielen}\ \emph {et~al.}(2018)\citenamefont {Gielen},
  \citenamefont {de~Le\'on~Ard\'on},\ and\ \citenamefont
  {Percacci}}]{Gielen2018}%
  \BibitemOpen
  \bibfield  {author} {\bibinfo {author} {\bibfnamefont {S.}~\bibnamefont
  {Gielen}}, \bibinfo {author} {\bibfnamefont {R.}~\bibnamefont
  {de~Le\'on~Ard\'on}}, \ and\ \bibinfo {author} {\bibfnamefont
  {R.}~\bibnamefont {Percacci}},\ }\href {\doibase 10.1088/1361-6382/aadbd1}
  {\bibfield  {journal} {\bibinfo  {journal} {Class. Quant. Grav.}\ }\textbf
  {\bibinfo {volume} {35}},\ \bibinfo {pages} {195009} (\bibinfo {year}
  {2018})},\ \Eprint {http://arxiv.org/abs/1805.11626} {arXiv:1805.11626
  [gr-qc]} \BibitemShut {NoStop}%
\bibitem [{\citenamefont {Wald}(1984)}]{Wald1984}%
  \BibitemOpen
  \bibfield  {author} {\bibinfo {author} {\bibfnamefont {R.~M.}\ \bibnamefont
  {Wald}},\ }\href {\doibase 10.7208/chicago/9780226870373.001.0001} {\emph
  {\bibinfo {title} {{General Relativity}}}}\ (\bibinfo  {publisher} {Chicago
  Univ. Pr.},\ \bibinfo {address} {Chicago, USA},\ \bibinfo {year}
  {1984})\BibitemShut {NoStop}%
\bibitem [{\citenamefont {Wigner}(1939)}]{Wigner1939}%
  \BibitemOpen
  \bibfield  {author} {\bibinfo {author} {\bibfnamefont {E.~P.}\ \bibnamefont
  {Wigner}},\ }\href {\doibase 10.2307/1968551} {\bibfield  {journal} {\bibinfo
   {journal} {Annals Math.}\ }\textbf {\bibinfo {volume} {40}},\ \bibinfo
  {pages} {149} (\bibinfo {year} {1939})}\BibitemShut {NoStop}%
\bibitem [{\citenamefont {Weinberg}(2005)}]{Weinberg1995}%
  \BibitemOpen
  \bibfield  {author} {\bibinfo {author} {\bibfnamefont {S.}~\bibnamefont
  {Weinberg}},\ }\href@noop {} {\emph {\bibinfo {title} {{The Quantum theory of
  fields. Vol. 1: Foundations}}}}\ (\bibinfo  {publisher} {Cambridge University
  Press},\ \bibinfo {year} {2005})\BibitemShut {NoStop}%
\bibitem [{\citenamefont {Fierz}\ and\ \citenamefont
  {Pauli}(1939)}]{Fierz1939}%
  \BibitemOpen
  \bibfield  {author} {\bibinfo {author} {\bibfnamefont {M.}~\bibnamefont
  {Fierz}}\ and\ \bibinfo {author} {\bibfnamefont {W.}~\bibnamefont {Pauli}},\
  }\href {\doibase 10.1098/rspa.1939.0140} {\bibfield  {journal} {\bibinfo
  {journal} {Proc. Roy. Soc. Lond. A}\ }\textbf {\bibinfo {volume} {173}},\
  \bibinfo {pages} {211} (\bibinfo {year} {1939})}\BibitemShut {NoStop}%
\bibitem [{\citenamefont {\'Alvarez}\ \emph {et~al.}(2006)\citenamefont
  {\'Alvarez}, \citenamefont {Blas}, \citenamefont {Garriga},\ and\
  \citenamefont {Verdaguer}}]{Alvarez2006}%
  \BibitemOpen
  \bibfield  {author} {\bibinfo {author} {\bibfnamefont {E.}~\bibnamefont
  {\'Alvarez}}, \bibinfo {author} {\bibfnamefont {D.}~\bibnamefont {Blas}},
  \bibinfo {author} {\bibfnamefont {J.}~\bibnamefont {Garriga}}, \ and\
  \bibinfo {author} {\bibfnamefont {E.}~\bibnamefont {Verdaguer}},\ }\href
  {\doibase 10.1016/j.nuclphysb.2006.08.003} {\bibfield  {journal} {\bibinfo
  {journal} {Nucl. Phys. B}\ }\textbf {\bibinfo {volume} {756}},\ \bibinfo
  {pages} {148} (\bibinfo {year} {2006})},\ \Eprint
  {http://arxiv.org/abs/hep-th/0606019} {arXiv:hep-th/0606019} \BibitemShut
  {NoStop}%
\bibitem [{\citenamefont {Bonifacio}\ \emph {et~al.}(2015)\citenamefont
  {Bonifacio}, \citenamefont {Ferreira},\ and\ \citenamefont
  {Hinterbichler}}]{Bonifacio2015}%
  \BibitemOpen
  \bibfield  {author} {\bibinfo {author} {\bibfnamefont {J.}~\bibnamefont
  {Bonifacio}}, \bibinfo {author} {\bibfnamefont {P.~G.}\ \bibnamefont
  {Ferreira}}, \ and\ \bibinfo {author} {\bibfnamefont {K.}~\bibnamefont
  {Hinterbichler}},\ }\href {\doibase 10.1103/PhysRevD.91.125008} {\bibfield
  {journal} {\bibinfo  {journal} {Phys. Rev. D}\ }\textbf {\bibinfo {volume}
  {91}},\ \bibinfo {pages} {125008} (\bibinfo {year} {2015})},\ \Eprint
  {http://arxiv.org/abs/1501.03159} {arXiv:1501.03159 [hep-th]} \BibitemShut
  {NoStop}%
\bibitem [{\citenamefont {Nakayama}(2023)}]{Nakayama2022}%
  \BibitemOpen
  \bibfield  {author} {\bibinfo {author} {\bibfnamefont {Y.}~\bibnamefont
  {Nakayama}},\ }\href {\doibase 10.1088/1361-6382/acd100} {\bibfield
  {journal} {\bibinfo  {journal} {Class. Quant. Grav.}\ }\textbf {\bibinfo
  {volume} {40}},\ \bibinfo {pages} {125005} (\bibinfo {year} {2023})},\
  \Eprint {http://arxiv.org/abs/2209.09462} {arXiv:2209.09462 [gr-qc]}
  \BibitemShut {NoStop}%
\bibitem [{\citenamefont {York}(1972)}]{York1972}%
  \BibitemOpen
  \bibfield  {author} {\bibinfo {author} {\bibfnamefont {J.~W.}\ \bibnamefont
  {York}, \bibfnamefont {Jr.}},\ }\href {\doibase 10.1103/PhysRevLett.28.1082}
  {\bibfield  {journal} {\bibinfo  {journal} {Phys. Rev. Lett.}\ }\textbf
  {\bibinfo {volume} {28}},\ \bibinfo {pages} {1082} (\bibinfo {year}
  {1972})}\BibitemShut {NoStop}%
\bibitem [{\citenamefont {Gibbons}\ and\ \citenamefont
  {Hawking}(1977)}]{GibbonsHawking1977}%
  \BibitemOpen
  \bibfield  {author} {\bibinfo {author} {\bibfnamefont {G.~W.}\ \bibnamefont
  {Gibbons}}\ and\ \bibinfo {author} {\bibfnamefont {S.~W.}\ \bibnamefont
  {Hawking}},\ }\href {\doibase 10.1103/PhysRevD.15.2752} {\bibfield  {journal}
  {\bibinfo  {journal} {Phys. Rev. D}\ }\textbf {\bibinfo {volume} {15}},\
  \bibinfo {pages} {2752} (\bibinfo {year} {1977})}\BibitemShut {NoStop}%
\bibitem [{\citenamefont {Beltr\'an~Jim\'enez}\ \emph
  {et~al.}(2018)\citenamefont {Beltr\'an~Jim\'enez}, \citenamefont
  {Heisenberg},\ and\ \citenamefont {Koivisto}}]{BeltranJimenez2018}%
  \BibitemOpen
  \bibfield  {author} {\bibinfo {author} {\bibfnamefont {J.}~\bibnamefont
  {Beltr\'an~Jim\'enez}}, \bibinfo {author} {\bibfnamefont {L.}~\bibnamefont
  {Heisenberg}}, \ and\ \bibinfo {author} {\bibfnamefont {T.~S.}\ \bibnamefont
  {Koivisto}},\ }\href {\doibase 10.1088/1475-7516/2018/08/039} {\bibfield
  {journal} {\bibinfo  {journal} {JCAP}\ }\textbf {\bibinfo {volume} {08}},\
  \bibinfo {pages} {039} (\bibinfo {year} {2018})},\ \Eprint
  {http://arxiv.org/abs/1803.10185} {arXiv:1803.10185 [gr-qc]} \BibitemShut
  {NoStop}%
\bibitem [{\citenamefont {Perez}\ and\ \citenamefont
  {Sudarsky}(2019)}]{Perez2017}%
  \BibitemOpen
  \bibfield  {author} {\bibinfo {author} {\bibfnamefont {A.}~\bibnamefont
  {Perez}}\ and\ \bibinfo {author} {\bibfnamefont {D.}~\bibnamefont
  {Sudarsky}},\ }\href {\doibase 10.1103/PhysRevLett.122.221302} {\bibfield
  {journal} {\bibinfo  {journal} {Phys. Rev. Lett.}\ }\textbf {\bibinfo
  {volume} {122}},\ \bibinfo {pages} {221302} (\bibinfo {year} {2019})},\
  \Eprint {http://arxiv.org/abs/1711.05183} {arXiv:1711.05183 [gr-qc]}
  \BibitemShut {NoStop}%
\bibitem [{\citenamefont {Alexander}\ and\ \citenamefont
  {Carballo-Rubio}(2020)}]{Alexander2018}%
  \BibitemOpen
  \bibfield  {author} {\bibinfo {author} {\bibfnamefont {S.}~\bibnamefont
  {Alexander}}\ and\ \bibinfo {author} {\bibfnamefont {R.}~\bibnamefont
  {Carballo-Rubio}},\ }\href {\doibase 10.1103/PhysRevD.101.024058} {\bibfield
  {journal} {\bibinfo  {journal} {Phys. Rev. D}\ }\textbf {\bibinfo {volume}
  {101}},\ \bibinfo {pages} {024058} (\bibinfo {year} {2020})},\ \Eprint
  {http://arxiv.org/abs/1810.02159} {arXiv:1810.02159 [gr-qc]} \BibitemShut
  {NoStop}%
\end{thebibliography}%

\end{document}